\documentclass[twocolumn, twocolappendix]{aastex631}

\usepackage{newtxtext,newtxmath,color}
\usepackage{amsmath, graphics, bm, graphicx, wasysym, verbatim}

\allowdisplaybreaks

\newcommand{\mathsym}[1]{{}}
\newcommand{\unicode}[1]{{}}

\newcommand{\ag}{\alpha}

\newcommand{\im}{{\rm i}}
\newcommand{\der}{{\rm d}}

\newcommand{\bxi}{{\bm \xi}}

\newcommand{\e}{{\rm e}}

\newcommand{\he}{{\bm {\hat e}}}

\newcommand{\del}{\nabla}
\newcommand{\br}{{\bm r}}

\newcommand{\Mjup}{M_{\rm Jup}}

\newcommand{\btimes}{{\bm \times}}
\newcommand{\bcdot}{{\bm \cdot}}
\newcommand{\bdel}{{\bm \del}}

\newcommand{\be}{\begin{equation}}
\newcommand{\ee}{\end{equation}}

\begin{document}

\title{Dynamical Tides during High-Eccentricity Migration produces\\the Hot Jupiter Pile-up, Neptune Ridge, and Neptune Desert}

\author[0000-0002-9849-5886]{J. J. Zanazzi}
\thanks{51 Pegasi b fellow\\email:jxz224@psu.edu}
\affiliation{
Department of Astronomy \& Astrophysics, 525 Davey Laboratory, The Pennsylvania State University, University Park, PA 16802, USA
}
\affiliation{
Center for Exoplanets and Habitable Worlds, 525 Davey Laboratory, The Pennsylvania State University, University Park, PA 16802, USA
}

\author[0000-0002-1417-8024]{Morgan MacLeod}
\affiliation{
Institute for Theory \& Computation, Center for Astrophysics, Harvard \& Smithsonian, Cambridge, MA 02138, USA
}

\author[0000-0002-6076-5967]{Marta L. Bryan}
\affiliation{
Department of Astronomy \& Astrophysics, 525 Davey Laboratory, The Pennsylvania State University, University Park, PA 16802, USA
}
\affiliation{
Center for Exoplanets and Habitable Worlds, 525 Davey Laboratory, The Pennsylvania State University, University Park, PA 16802, USA
}

\author[0000-0001-9596-7983]{Suvrath Mahadevan}
\affiliation{
Department of Astronomy \& Astrophysics, 525 Davey Laboratory, The Pennsylvania State University, University Park, PA 16802, USA
}
\affiliation{
Center for Exoplanets and Habitable Worlds, 525 Davey Laboratory, The Pennsylvania State University, University Park, PA 16802, USA
}




\begin{abstract}
The period distribution of hot gaseous exoplanets depends strongly on mass. Clustering between the orbital periods of $\sim$3 to $\sim$5-6 days is seen for sub-Saturns (``Neptune ridge'') and Jovians (``hot Jupiter pile-up''), contrasting with a sharp deficit interior to 3 days for sub-Saturns, not seen for Jovians (``Neptune desert''). During high-eccentricity migration, tidally-excited fundamental-modes ($f$-modes) act as a reservoir for orbital energy, and can take gaseous planets formed beyond several AU and place them at short separations. However, how $f$-modes relinquish their energy into the planet interior is unknown. Here, we show how $f$-modes can not only circularize orbits --- causing clustering near the Neptune ridge and hot-Jupiter pile-up --- but can also shock and unbind mass, leaving sub-Saturn cores in the Neptune desert. Our hydrodynamical simulations demonstrate that close approaches tidally excite $f$-modes, whose super-sonic velocities shock gaseous envelopes. Atmospheres cool by radiative diffusion or winds when shocks penetrate shallow versus deep depths. Planetary structural and orbital evolution is followed over many periastron passages using an iterative map: shocks that diffusively cool circularize and bunch orbits near the hot Jupiter pile-up and Neptune ridge, while shocks that drive outflows unbind envelopes and place gas giant cores in the Neptune desert. Sub-Saturns that dwell in the desert are predicted to arrive with large spin-orbit misalignments after producing luminous flares.
\end{abstract}



\section{Introduction}

The observed properties of gas giants at sub-ten-day orbital periods, or hot Jupiters, suggest that they had raucous pasts. Massive envelopes require seed $\sim 10 M_\oplus$ cores to accrete \citep[e.g.][]{Pollack+1996}, that grow only beyond the ice sublimation radius. Jovians that suffer strong interactions with companion planets or stars can be placed on highly eccentric and misaligned orbits, and attain short separations after migrating due to the dissipation of the tidal bulge raised on the planet by the host. The observed misalignments between the host equator and planet orbit, and preference for $\sim$3-5 day orbital periods, are expected outcomes if hot Jupiters form via high-eccentricity migration \citep[see][for review]{DawsonJohnson2018}.

The demographics of planets with masses below that of Saturn differ markedly from hot Jupiters at the shortest separations. At sub-three-day orbital periods, planets with masses between $10 M_\oplus$ to $100 M_\oplus$ are scarce, a demographic feature called the ``sub-Saturn desert'' or ``Neptune desert'' \citep[e.g.][]{SzaboKiss2011, BeaugeNesvorny2013, Mazeh+2016, Lundkvist+2016}. Although originally hypothesized to stem from hydrodynamical winds driven by the host star's X-ray and Extreme-Ultraviolet (XUV) irradiation \citep[e.g.][]{Jackson+2012, KurokawaNakamoto2014}, photoevaporation alone is insufficient to strip the envelopes of $\gtrsim 20 M_\oplus$ sub-Saturns at short separations \citep[e.g.][]{OwenLai2018, Vissapragada+2022, Saidel+2025}. Instead, high-eccentricity migration is thought to play a role. 
Following eccentricity excitation after forming far from their hosts, because orbital angular momentum is roughly conserved as planets circularize, hot Jupiters arrive at semi-major axes roughly twice their initial pericenter distances. Tidal disruption when pericenter passages lie interior to the Roche limit predict an inner boundary for Jovians and sub-Saturns, which lies in agreement with the Neptune desert edge \citep[e.g.][]{FordRasio2006, MatsakosKonigl2016, OwenLai2018, Castro-Gonzalez+2026}.
Sub-Saturns clumped directly exterior to the desert within the orbital periods of 3-6 days (``Neptune ridge'') tend to have large spin-orbit misalignments, lending support to high-eccentricity migration playing a role in placing hot Neptunes at their observed locations \citep[e.g.][]{Bourrier+2023, Castro-Gonzalez2024a, Yee+2025}.

\begin{figure}
  \centering
  \includegraphics[width=\columnwidth]{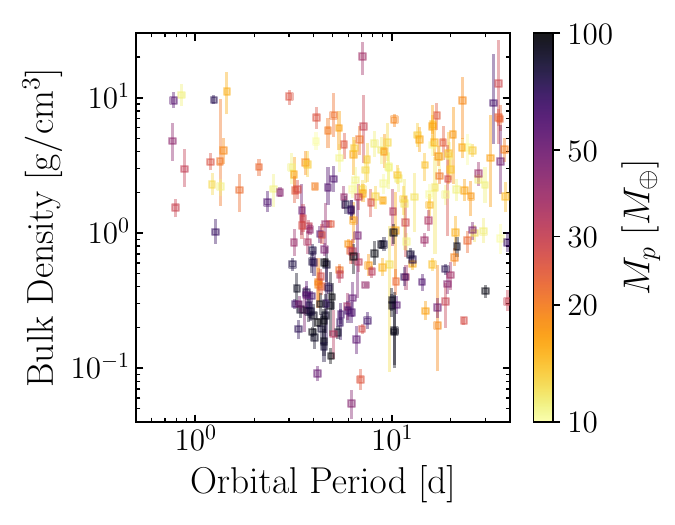}
  \caption{Sub-Saturns in the Neptune desert (masses $10 M_\oplus$ to $100 M_\oplus$, $<$3 day orbital periods) are more dense than their longer-period counterparts. Bulk densities and 1$\sigma$ errors are plotted with orbital period, with colors denoting the planet masses indicated. Mass, radius, and orbital period measurements for sub-Saturns with F/G/K hosts ($0.6 M_\odot$ to $1.4 M_\odot$) are from the NASA Exoplanet Archive.
  }
  \label{fig:subSat_density}
\end{figure}

Although bare, the Neptune desert is not barren, as some sub-Saturns have eked their way to sub-three-day periods \citep[e.g.][]{Jenkins+2020, Armstrong+2020, Naponiello+2023, Osborn+2023, Nabbie+2024, Doyle+2025, Carleo+2026}. The few dozen detected ``desert dwellers'' or ``desert Neptunes'' are much more compact than their less-irradiated counterparts, with bulk densities between $\sim$1-10~${\rm g}/{\rm cm}^3$ (Fig.~\ref{fig:subSat_density}, see also e.g.~\citealt{Castro-Gonzalez+2024b, Doyle+2025}). Interestingly, \cite{VissapragadaBehmard2025} showed the host stars of desert Neptunes and hot Jupiters are similarly metal rich, and posit that desert dwellers are exposed cores of failed hot Jupiters.

Despite the substantial literature on high-eccentricity migration, scant attention has been paid into how heat deposited during circularization might affect planetary structure. Simple energetics
suggest the effect could be dramatic, since circularization deposits $\sim G M_\odot M_{\rm Jup}/{\rm AU} \sim 10^{43}$~erg of orbital energy into the hot Jupiter, whose binding energy $\sim G M_{\rm Jup}^2/R_{\rm Jup} \sim 10^{43}$~erg. 
If circularization rapidly deposits energy at deep depths, the internal structure of the planet could be significantly affected \citep[e.g.][]{Podsiadlowski1996, Leconte+2010, Millholland+2020, Glanz+2022, HallattMillholland2026b}.
Recent works show that energies of fundamental-modes ($f$-modes) excited during periastron passages can become comparable to the binding energy \citep[e.g.][]{VickLai2018, Wu2018, Yu+2021, Yu+2022}, but how $f$-mode energy is relinquished to the planet interior remains unclear. Hydrodynamical simulations have demonstrated that successive periastron passages can unbind gas giant envelopes \citep{Faber+2005, Guillochon+2011, Liu+2013, FanLiu2026}. However, these works focused more on the direct tidal stripping of gas, and less on the heat deposited by excited $f$-modes.

This work investigates how the damping of seismic oscillations can circularize orbits and drive outflows that deplete a giant of its gas. 
Section~\ref{sec:hydro_sim} demonstrates how gas giants respond tidally to close periastron passages with their host stars, using a suite of hydrodynamical simulations. Section~\ref{sec:Outcomes} investigates how circularization from dynamical tides can affect the occurrence rates of hot Jupiters and Neptunes, using an iterative map to chart the structural and orbital evolution of planets with $f$-modes excited after close approaches. Section~\ref{sec:SummDisc} summarizes our results, and discusses their implications.

\section{Wave Breaking Heats Atmospheres and Drives Outflows}
\label{sec:hydro_sim}

During a gas giant periastron passage, tidal forcing from the host star excites surface gravity waves. Here our hydrodynamical simulations demonstrate that close encounters elicit flows so large that their velocities become supersonic near the surface and shock. We find shocks not only damp $f$-mode amplitudes, causing rapid circularization during high-eccentricity migration, but can also drive outflows. We describe our simulation setup in Section~\ref{sec:setup}, present our results in Section~\ref{sec:sim_results}, and compare to normal mode theory expectations in Section~\ref{sec:CompLT}.

\subsection{Hydrodynamical Simulations Setup}
\label{sec:setup}

We aim to capture heating in the regions near the planet surface. For such low-density environments, volume-resolving Eulerian grid-based codes are preferred to Lagrangian mass-based methods. Previous works have shown that the grid-based \texttt{Athena++} code \citep{Stone+2020} can successfully capture the tidal interaction between a star and a companion \citep[e.g.][]{MacLeod+2019, MacLeod+2022, MacLeodLoeb2023}, and a rocky planet with an M-dwarf host \citep{LoebMacLeod2024}. Here, we discuss the pertinent aspects of our simulation setup. We refer the reader to \cite{MacLeod+2018, MacLeod+2019} for further details on the code that we use.

We simulate a planet with mass $M_p$, radius $R_p$, on an eccentric orbit around a host star with mass $M_\star$ and position ${\bm r}_\star$ relative to the planet. Our simulation units set the planet mass, radius, and Newton's gravitational constant to unity ($G = M_p = R_p = 1$), so that the units of velocity $v_p = (G M_p/R_p)^{1/2}$, time $t_p  = (G M_p/R_p^3)^{-1/2}$, density $M_p/R_p^3$, and energy $E_p = G M_p^2/R_p$. The mass-ratio $M_\star/M_p = 10^3$ and semi-major axis $a = 10^3 R_p$ ($0.48$~AU for $R_p = R_{\rm Jup}$) are fixed, varying the orbital eccentricity $e$ so that the pericenter distance $r_{\rm peri} = a(1-e)$ takes the values $\{1.8, 2.0, 2.2\} r_t$, where $r_t = R_p (M_\star/M_p)^{1/3}$ is the tidal radius. 

\begin{figure}
  \centering
  \includegraphics[width=\columnwidth]{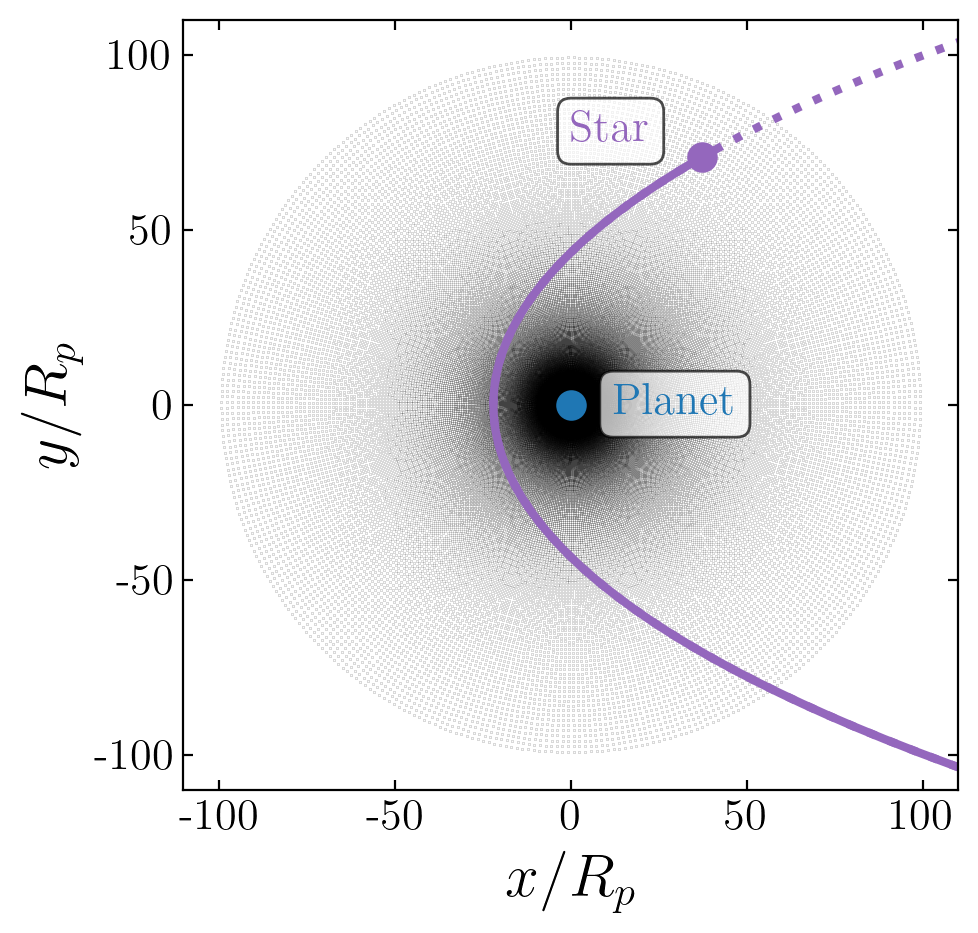}
  \caption{Orbit of star in spherical, planet-centered frame. Grid points are displayed with black circles, the star's starting position at $r=80 R_p$ with the purple dot, and the star's orbit over the course of the simulation with the solid purple line. The simulation continues until the star exceeds the distance $r = 480 R_p$ at time $t=200 t_p$ (both numbers arbitrary, with final position outside plot). We view in the direction of the orbit normal, for a pericenter distance $r_p = 2.2 r_t$.
  }
  \label{fig:HJ_peri_pass}
\end{figure}

To resolve the shock-heating of the atmosphere, we work in a spherical coordinate system co-moving with the planet (Fig.~\ref{fig:HJ_peri_pass}). The base spherical mesh $(r,\theta,\varphi)$ has 480$\times$192$\times$384 zones, spaced logarithmically in $r$ from $0.1 R_p$ to $100 R_p$, and uniformly in $\theta$, $\varphi$ over the full $4\pi$ steradians of solid angle. Between $0.7 R_p$ and $1.3 R_p$, a nested mesh increases the radial grid resolution by a factor of two. This spatial resolution is chosen to most-closely resolve the region near the limb of the planet where pressure gradients are steepest. We find that this is the spatial resolution needed to ensure that numerical hydrostatic balance is well maintained for the duration of the numerical calculation, such that departures from the original structure are due to tidal forcing \citep{MacLeod+2018,MacLeod+2022}. Performing calculations with several periapse distances allows us to verify this by comparing weaker and stronger tidal passages.   Boundary conditions in $\theta$ allow material to flow through the coordinate singularities at $0$ and $\pi$, $\varphi$ boundaries are periodic. We adopt a reflecting boundary condition at $r = 0.1 R_p$, and an outflow (but no inflow) boundary condition at $r = 100 R_p$. 

The planet mass is broken into two pieces $M = M_c + M_g$, where $M_c = 0.1 M_p$ is the constant mass of the core interior to $r<0.1 R_p$, while $M_g$ is the mass of gas in the \texttt{Athena++} simulation. The initial pressure and density profile is initialized by placing material exterior to $M_c$, and solving the equations of hydrostatic equilibrium for an adiabatic equation of state $P \propto \rho^\Gamma$, setting $\Gamma = 5/3$ for an ideal gas. We choose the central pressure and density so that the total planet mass is initially $M = M_p = 1$, and $\rho$ vanishes at $r = R_p$. Exterior to $R_p$, the planet is connected to a continuous, artificial, low-density background in hydrostatic equilibrium with the planetary gravity with constant sound speed $v_p/3$, that declines from $\sim 10^{-5} M_p/R_p^3$ near $R_p$ to $\sim 10^{-9} M_p/R_p^3$ near $100 R_p$. We assign scalar tracers to material originating from the planet envelope, to distinguish from the low-density isothermal background. The simulation is initialized at $t=0$ with a planet-star distance of $70 R_p$ before the periastron passage time $t_{\rm peri}$, and runs until $t = 200 t_p$. 

As in \cite{MacLeod+2019, MacLeod+2022, MacLeodLoeb2023}, we express the continuity, momentum, and energy equations in the frame co-moving with the planet:
\begin{align}
    &\partial_t \rho + \bdel \bcdot (\rho {\bm v}) = 0,
    \label{eq:cont} \\
    &\partial_t(\rho {\bm v}) + \bdel \bcdot (\rho {\bm v} {\bm v} + P {\bf I}) = -\rho {\bm a}_{\rm ext}, \\
    &\partial_t E + \bdel \bcdot \big[(E + P){\bm v} \big] = - \rho {\bm v} \bcdot {\bm a}_{\rm ext}.
\end{align}
The planet density $\rho$, velocity ${\bm v}$, and energy $E = \epsilon + \rho {\bm v} \bcdot {\bm v}/2$ are solved at each timestep, assuming an ideal gas equation of state $P = (\Gamma - 1) \epsilon = \frac{2}{3} \epsilon$. The external acceleration is comprised of multiple terms ${\bm a}_{\rm ext} = {\bm a}_\star + {\bm a}_p - {\bm a}_{p, {\rm acc}}$ : the tidal acceleration from the host star ${\bm a}_\star \approx -G M_\star ({\bm r} - {\bm r}_\star) f_{\rm spline}(|{\bm r} - {\bm r}_\star|, r_{\rm soft})$ (with $f_{\rm spline}$ being eq. A2 from \citealt{HernquistKatz1989}), the planet's self-gravity implemented through the monopole approximation, ${\bm a}_p = -G M(<r){\bm r}/r^3$, subtracting planet's acceleration, ${\bm a}_{p, {\rm acc}} = G M_\star {\bm r}_\star/r_\star^3 + \int_{\rm gas} (G \rho {\bm r}/r^3){\rm d}V$. We soften the star's acceleration by $r_{\rm soft} = 0.1 R_p$. See \cite{MacLeod+2018} for further details on the implementation of ${\bm a}_{\rm ext}$. We note in the monopole approximation, the planet's self-gravity ${\bm a}_p$ is treated as spherically symmetric, neglecting any deviations in the gravitational potential from fluid flows, analogous to the Cowling approximation for normal modes \citep{Cowling1941}. For $f$-modes excited by close tidal encounters, self-gravity can be an up to order-unity correction to the oscillation frequencies and amplitudes.

\subsection{Hydrodynamical Simulation Results}
\label{sec:sim_results}

\begin{figure*}
  \centering
  \includegraphics[width=\linewidth]{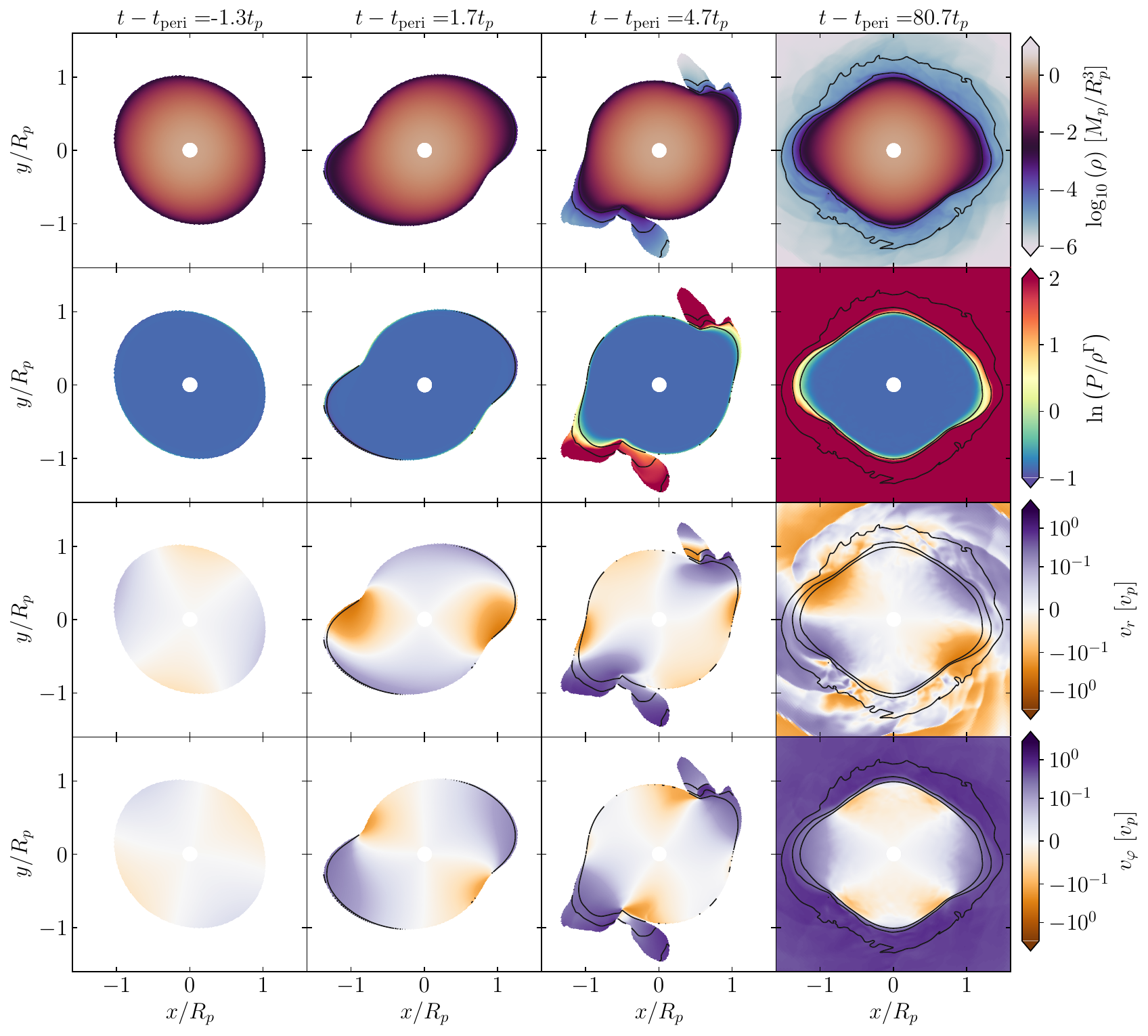}
  \caption{The pericenter passage excites tidal flows that shock the planet surface. We view the spatial dependence $(x,y)$ normal to the orbital plane for the density $\rho$ (top), logarithm of entropy $\ln (P/\rho^{\Gamma})$ (second), radial velocity $v_r$ (third), and azimuthal velocity $v_\varphi$ (bottom  row), at times relative to the time of closest approach $t_{\rm peri}$ displayed by columns. Contours delineate densities of $10^{-5}, 10^{-4}, 10^{-3} M_p/R_p^3$. Panels display white when the scalar tracer has a density below $0.9$, when at least $10\%$ of the low-density background is mixed within the planetary flow. Here, a pericenter distance of $r_{\rm peri} = 2.2 r_t$ is displayed.
  }
  \label{fig:peri_pass_profiles}
\end{figure*}

\begin{figure*}
  \centering
  \includegraphics[width=\linewidth]{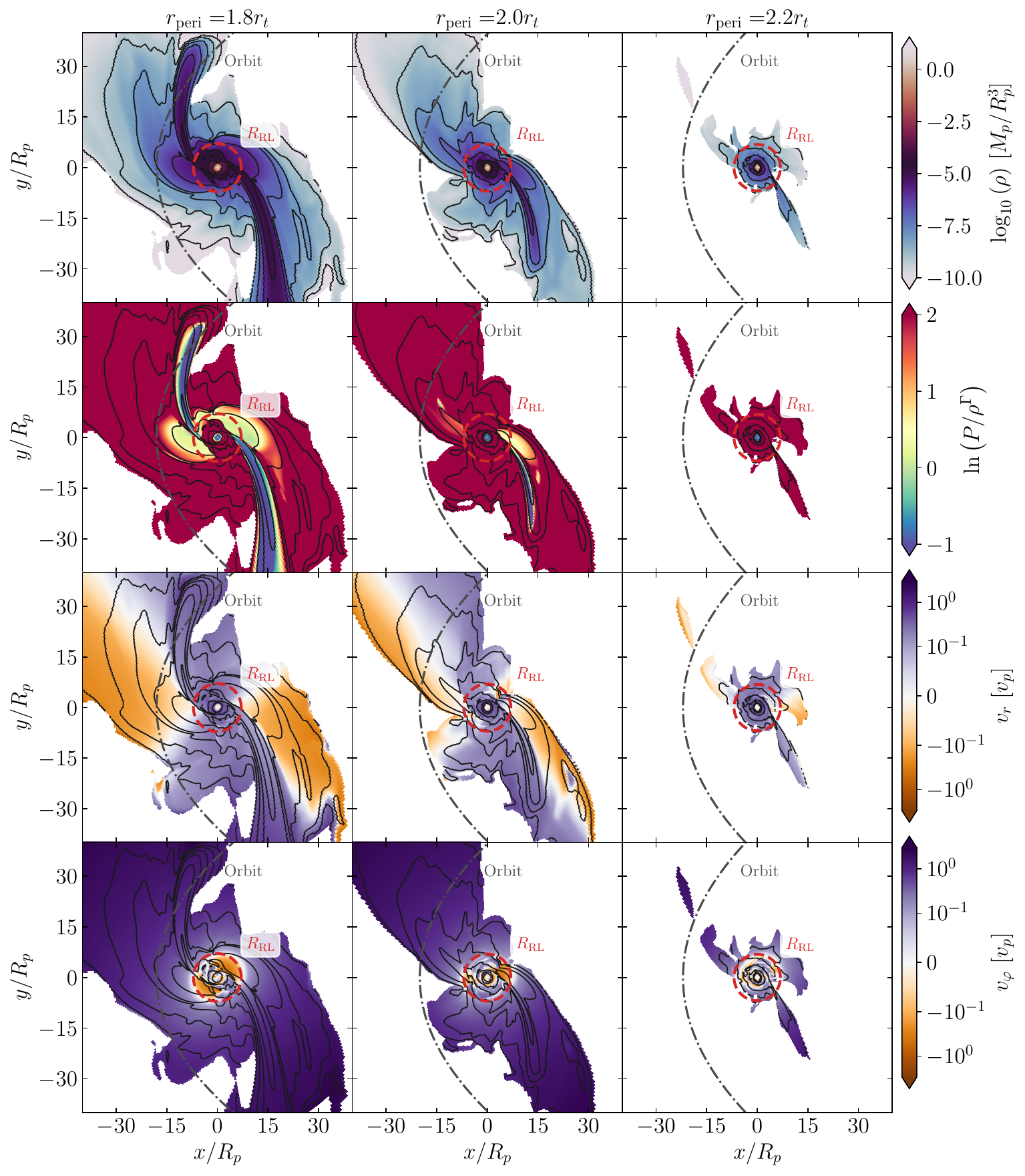}
  \caption{Tidal shocks drive an outflow from the planet surface, that extends tens of $R_p$. Panels display orbits with the $r_{\rm peri}$ values indicated, while colors display density (top), entropy (second), radial velocities (third), and azimuthal velocities (bottom) from our simulations, with black contours denoting $10^{-10}, 10^{-9}, 10^{-8}, \dots, 10^{-3} M_p/R_p^3$ densities. White is displayed when the scalar tracer has a density below $0.5$, or most material is from the low-density background. Dashed red line denotes the Roche lobe $R_{\rm RL}$, while gray dot-dashed line displays the host orbit in the planet-centered frame. Results displayed are evaluated at the time $t-t_{\rm peri} = 31 t_p$, viewed parallel to the orbit normal.
  }
  \label{fig:peri_pass_outflow}
\end{figure*}

Similar to \cite{MacLeod+2022}, we find that close pericenter passages shock the atmosphere, creating a diffuse envelope. Before periastron ($t < t_{\rm peri}$), tidal flows are well described by linear normal modes (Fig.~\ref{fig:peri_pass_profiles} left panel). Radial and azimuthal velocities $v_r$ and $v_\varphi$ approximate $m=2$ displacements from $\ell=2$ $f$-mode oscillations. The body is roughly isentropic, and density remains unperturbed. Soon after periastron ($0 <t - t_{\rm peri} \lesssim 3 t_p$), flows steepen as inward radial velocities exceed the planet sound-speed (Fig.~\ref{fig:peri_pass_profiles} second panel). Velocities no longer resemble $m=2$ displacements from linear normal modes; higher harmonics in $m$ and $\ell$ are visually apparent. The planet surface becomes visibly deformed. Eventually, tidal flows shock at the surface (Fig.~\ref{fig:peri_pass_profiles} third panel). Radial and azimuthal exceed the escape velocity $v_{\rm esc} = 2^{1/2} v_p$, ejecting material. The entropy spikes on opposite sides of the planet, heating outflows. Later, shocks heat a low-density envelope that surround the planet ($t - t_{\rm peri} \gtrsim 10 t_p$, Fig.~\ref{fig:peri_pass_profiles}). The diffuse envelope has high entropy due to heating from shocks. Angular momentum deposited by the flow causes the envelope to rotate near the Keplerian angular frequency $\sim v_p/R_p$. Non-spherical, radial flows emanate from the planet surface.

At late times, shock-heated flows expel mass (Fig.~\ref{fig:peri_pass_outflow}). Interior to the Roche radius $R_{\rm RL} \approx 0.49 (M_p/M_\star)^{1/3} r_\star$, the outflow is roughly spherically-symmetric. The density slowly tapers off, and has a high entropy. Radial velocities are positive, indicating shock-heating uniformly expels mass. A few $R_p$ from the surface, but interior to $R_{\rm RL}$, the bound envelope has little angular momentum, and azimuthal velocities are low. 

Exterior to $R_{\rm RL}$, tides significantly modify the outflow morphology. The density still tapers off with distance, but material is concentrated into two diametrically-opposed tidal tails. Because mass within the tails is stripped by the tidal encounter, rather than expelled by shock-heating, material is dense, cold, and follows ballistic orbits. Entropy is high outside the tails, heated by the diffuse hot gas from the simulation initial conditions. The mass entrained in the cool ($\ln[P/\rho^\Gamma]<0.5$) tails exterior to $r>1.5 R_p$ is $1.6 \times 10^{-4}$, $1.2 \times 10^{-6}$ and $0$ $M_p$ for $r_{\rm peri} = 1.8, 2.0$, and $2.2 r_t$, respectively, much smaller than the mass of the high-entropy material interior to $R_{\rm RL}$ (see Fig.~\ref{fig:analytic_vs_sim_scaling} middle panel, below). Radial velocities are largest in the tidal tails. Azimuthal velocities outside of $R_{\rm RL}$ are high because of tidal shearing. Deeper periastron passages generally drive stronger outflows, expelling more mass to larger distances from the planet. Material past $R_{\rm RL}$ is poorly resolved for our weakest encounter ($r_{\rm peri} = 2.2 r_t$), because its density becomes similar to the numerical background.

Simulating wider periastron passages, we find tidal flows exhibit negligible shocks that are unable to drive outflows (Appendix~\ref{sec:wide_peri}). Numerically, we find shocks occur at pericenters $r_{\rm peri} \lesssim 2.4 r_t$, and outflows are driven when $r_{\rm peri} \lesssim 2.2 r_t$. The conditions needed for tidal flows to shock are discussed more in the next section.

\subsection{Simulation Scaling using Normal Mode Theory}
\label{sec:CompLT}

Before the surface gravity waves shock the planet surface, velocity amplitudes are small and describable by the superposition of normal modes. Not only can normal mode theory reproduce simulated quadrupolar flow amplitudes, but can also estimate the mass within the region surface gravity waves shock, and the mode damping rate from wave breaking \citep{MacLeod+2022}. These oscillation scalings will be utilized in our calculations that track the long-term structure and orbital evolution of gaseous planets. Although we adopt the nearly parabolic encounter model of e.g. \cite{Lai1997, IvanovPapaloizou2004, VickLai2018, Wu2018, Vick+2019}, we use the notation and normalizations of e.g. \cite{Weinberg+2012, Yu+2021, Yu+2022}, and assume $M_p \ll M_\star$.

We consider a normal mode with displacement $\bxi(t,\br)$ in a frame co-rotating with the planet angular frequency $\Omega_p$, and describe it through phase-space expansion of normal modes $\tilde \bxi_\alpha(\br)$, with eigenfrequencies $\omega_\alpha$ and amplitudes $q_\alpha(t)$
\begin{equation}
    \left[
    \begin{array}{c}
    \bxi(t,\br) \\
    \partial_t \bxi(t,\br)
    \end{array} \right]= \sum_\alpha q_\alpha(t)
    \left[ \begin{array}{c}
    \tilde \bxi_\alpha(\br) \\
    -\im \omega_\alpha \tilde \bxi_\ag(\br)
    \end{array} \right]{}\e^{-\im \omega_\alpha t}.
    \label{eq:bxi_exp}
\end{equation}
Each eigenmode $\tilde \bxi_\alpha(\br)$ satisfies
\begin{equation}
    - \omega_\alpha^2 \tilde \bxi_\alpha + 2 \im \omega_\alpha {\bm \Omega}_p \btimes \tilde \bxi_\alpha + \mathcal{L} \bcdot \tilde \bxi_\alpha = 0,
    \label{eq:eigenvector}
\end{equation}
where the linear operator $\mathcal{L}$ takes into account the restoring force from pressure, buoyancy, and self-gravity on the oscillation. For rotating planets, distinct eigenvectors $\tilde \bxi_\alpha$, $\tilde \bxi_{\alpha'}$ are not orthogonal under the inner product $\langle \tilde \bxi_\ag| \tilde \bxi_{\ag'} \rangle = \int \tilde \bxi_\ag^* \bcdot \tilde \bxi_{\ag'} \rho \der V$, where $X^*$ denotes the complex-conjugate of $X$, which complicates the calculation of $q_\alpha(t)$. However, \cite{Schenk+2001} showed that when linear normal modes feel an external driving force $-\bdel U_{\rm ext}$, $q_\alpha$ satisfy
\begin{equation}
    \dot q_\alpha + \im \omega_\alpha q_\alpha = \langle \tilde \bxi_\alpha | -\bdel U_{\rm ext} \rangle.
    \label{eq:dqdt_schenk+2001}
\end{equation}
Here, normalized mode energies in the rotating frame are given by
\begin{equation}
    \tilde \varepsilon_\alpha = 2 \omega_\alpha^2 \langle \tilde \bxi_\alpha | \tilde \bxi_\alpha \rangle + 2 \omega_\alpha \langle \tilde \bxi_\alpha | {\bm \Omega}_p \btimes \tilde \bxi_\alpha \rangle,
\end{equation}
where we have added mode $\tilde \bxi_\alpha \e^{-\im \omega_\alpha t}$ to its physically-identical complex conjugate $\tilde \bxi_\alpha^* \e^{+\im \omega_\alpha t}$. Derivation of equation~\eqref{eq:dqdt_schenk+2001} involves manipulating the Jordan chain created by the phase space mode expansion $\{\tilde \bxi_\alpha(\br), -\im \omega_\alpha \tilde \bxi_\alpha(\br)\}$ \citep{Schutz1979, DysonSchutz1979, Schutz1980a, Schutz1980b}. We neglect how Coriolis forces affect normal modes, effectively setting $\Omega_p = 0$ in our calculations, because our simulations initialize the planet as non-rotating. However, we use equation~\eqref{eq:bxi_exp} in our calculation of $q_\alpha(t)$, so that our results are easily generalizable to include rotation, which can affect the Love numbers \citep[e.g.][]{Lai2021, DewberryLai2022} and dissipation rates \citep[e.g.][]{DewberryWu2024, Fuller+2026} of giant planet $f$-modes. For the rest of this paper, we drop $\alpha$ subscripts, and normalize $\tilde \bxi$ so that $\tilde \varepsilon = E_p$.

Equation~\eqref{eq:bxi_exp} allows us to calculate the change in mode amplitude after a periastron passage. Because surface gravity waves have frequencies $\omega \gtrsim (G M_p/R_p)^{1/2} \gg \Omega_p$ \citep[e.g.][]{Zanazzi+2025}, we neglect rotation and set $\Omega_p = 0$. Equation~\eqref{eq:eigenvector} is then separable, with $\tilde \bxi(r,\theta,\varphi)$ expressible in terms of vector spherical harmonics:
\begin{equation}
    \tilde \bxi(r,\theta,\varphi) = \tilde \xi_r(r) Y_{\ell m}(\theta, \varphi) \he_r + \tilde \xi_\perp(r) r \bdel Y_{\ell m}(\theta, \varphi).
    \label{eq:tbxi_sph}
\end{equation}
For arbitrary $U_{\rm ext}$, equation~\eqref{eq:dqdt_schenk+2001} needs to be summed over all angular degrees $\ell$ and azimuthal numbers $m$. But because the tidal potential from the host acting on the planet can also be expressed as a sum over spherical harmonics \citep{PressTeukolsky1977}
\begin{align}
    &U_{\rm ext} = \sum_{\ell m} U_{\ell m} 
    \nonumber \\
    &= \sum_{\ell = 2}^{\infty} \sum_{m=-\ell}^\ell W_{\ell m} \frac{G M_\star r^\ell}{D(t)^{\ell+1}} \e^{ -\im m \Phi(t) } Y_{\ell m}(\theta,\varphi),
\end{align}
each tidal potential linearly forces an eigenmode with unique $\{\ell, m\}$, so that
\begin{equation}
    \dot q + \im \omega q = \im \omega \tilde U_{\ell m},
    \label{eq:dqdt_lm}
\end{equation}
where
\begin{align}
    &\tilde U_{\ell m} = W_{\ell m} \tilde I_{\ell m} \frac{M_\star}{M_p} \left[ \frac{R_p}{D(t)} \right]^{\ell+1} \e^{ -\im m \Phi(t)  },\\
    &W_{\ell m} = (-)^{(\ell+m)/2} \left[ \frac{4\pi}{2\ell+1} (\ell+1)!(\ell-1)!\right]^{1/2}
    \nonumber \\
    &\hspace{20mm} \times \left[ 2^\ell \left( \frac{\ell-m}{2} \right)! \left( \frac{\ell+m}{2} \right)!  \right]^{-1},\\
    &\tilde I_{\ell m} = \frac{1}{M_p R_p^\ell} \Big\langle \tilde \bxi | \bdel\big(r^\ell Y_{\ell m}\big) \Big\rangle
    \label{eq:I_lm}
\end{align}
with $(-)^k$ having values of $(-1)^k$ for integer $k$, and $0$ otherwise, while $D(t)$ is the host-planet distance, and $\Phi(t)$ the true anomaly. Equation~\eqref{eq:dqdt_lm} can be integrated to give the change in $q$ after one orbit:
\begin{equation}
    \Delta q = \im 2\pi t_p \omega \tilde I_{\ell m} K_{\ell m} \frac{M_\star}{M_p} \left( \frac{R_p}{r_{\rm peri}} \right)^{\ell+1},
    \label{eq:Dq}
\end{equation}
where
\begin{equation}
    K_{\ell m} = \frac{W_{\ell m}}{2\pi t_p} \int\limits_{-P_{\rm orb}/2}^{P_{\rm orb}/2} \left[ \frac{r_{\rm peri}}{D(\tau)} \right]^{\ell+1} \e^{\im [\omega \tau - m \Phi(\tau)]} \der \tau.
    \label{eq:K_lm}
\end{equation}

\begin{figure*}[htbp]
  \centering
  \includegraphics[width=\textwidth]{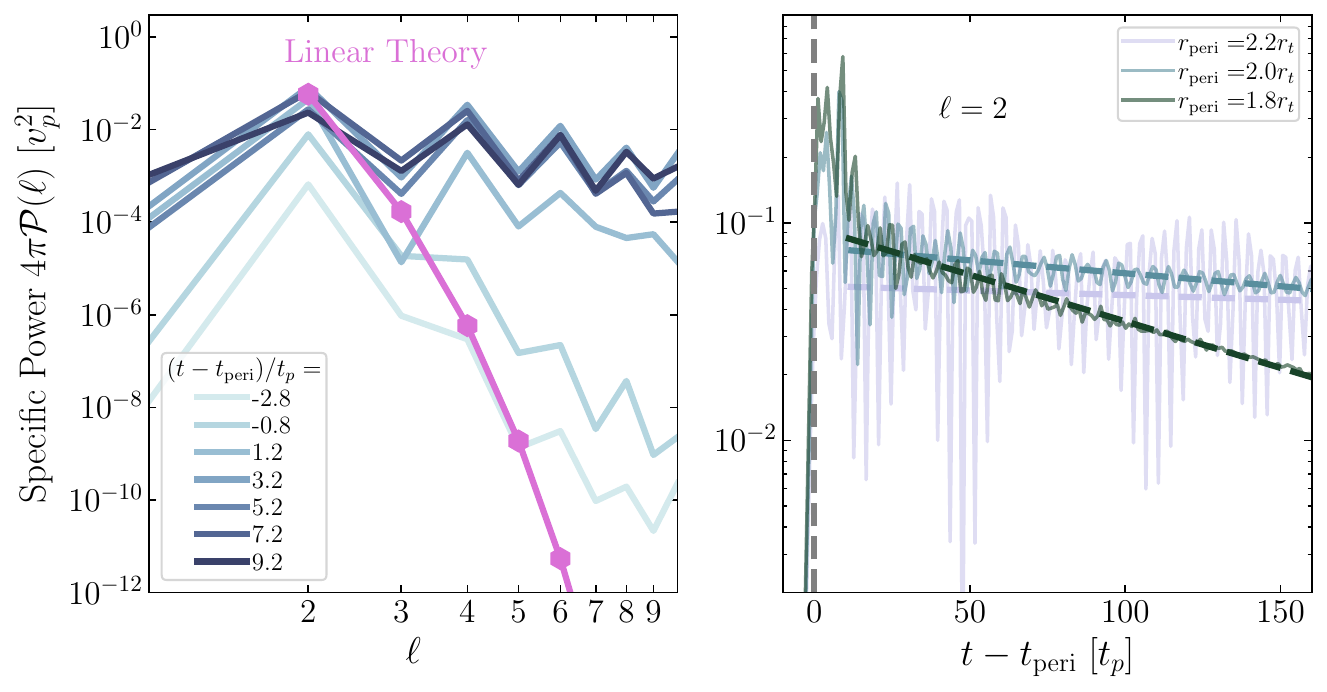}
  \caption{Surface radial velocity power spectrum (eq.~\ref{eq:PowerSpec}). \textit{Left panel}: Power dependence on $\ell$ is well-described by linear theory before periastron $t_{\rm peri}$, and deviates after because of turbulence excited by wave breaking. We display angular degree $\ell$ at times $t$ indicated, with $r_{\rm peri} = 2.2 r_t$. \textit{Right panel}: The exponential decay of the $\ell=2$ power with time. Thin opaque lines show the simulation power for the $r_{\rm peri}$ values indicated, while dashed lines display the exponential fits to the simulations $\mathcal{P}(2) \propto \e^{-(t-t_{\rm peri})\gamma_{\rm wb}}$.
  }
  \label{fig:power_spectrum}
\end{figure*}

\begin{figure}
  \centering
  \includegraphics[width=\linewidth]{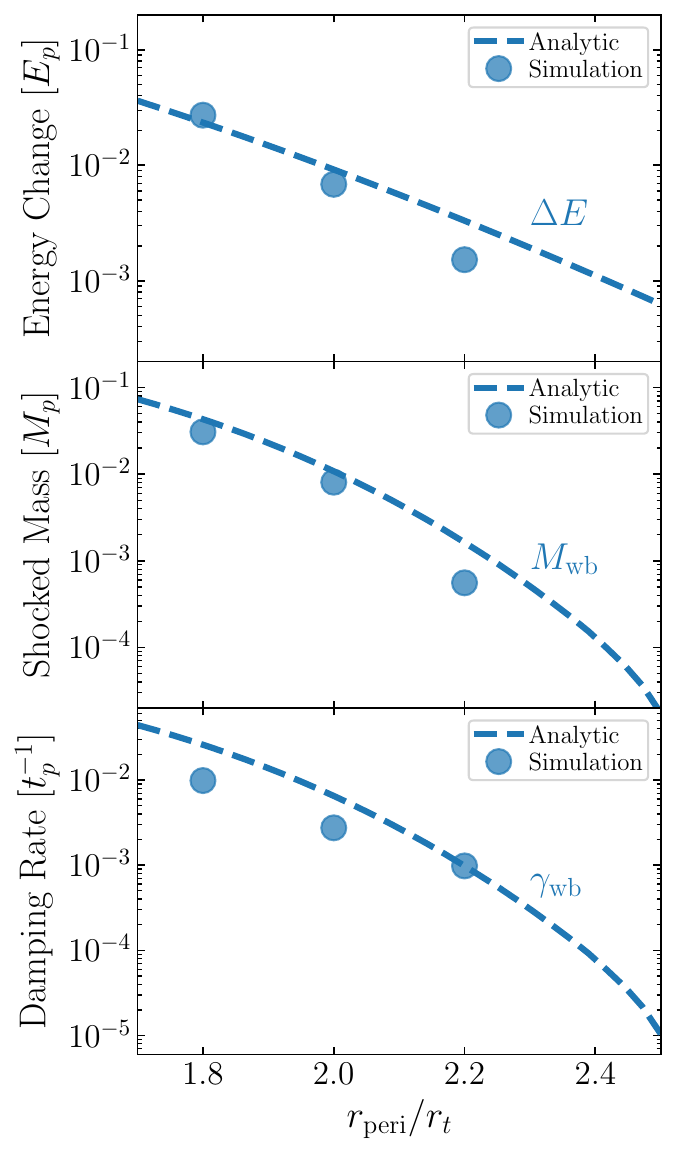}
  \caption{Our hydrodynamical simulations (dots) agree with normal mode expectations (dashed lines), for the $r_{\rm peri}$ values considered. \textit{Top panel}: Total change in kinetic, potential, and internal energy of the planet at the end of the simulation, compared to the energy imparted into the $f$-mode after passage through periastron (eq.~\ref{eq:DE_lin}). \textit{Middle panel}: Total mass of high entropy material in our simulations ($\ln[P/\rho^\Gamma]>0$), compared to the mass enclosed in the region where the $f$-mode is supersonic (eq.~\ref{eq:M_wb}). \textit{Bottom panel}: Damping rate of the simulated $\ell=2$ power (Fig.~\ref{fig:power_spectrum} right panel), compared to the predicted damping rate from wave breaking (eq.~\ref{eq:gamma_wb}). We set the order-unity coefficient $\epsilon_{\rm wb} = 0.3$. 
  }
  \label{fig:analytic_vs_sim_scaling}
\end{figure}

\begin{figure}[htbp]
  \centering
  \includegraphics[width=\linewidth]{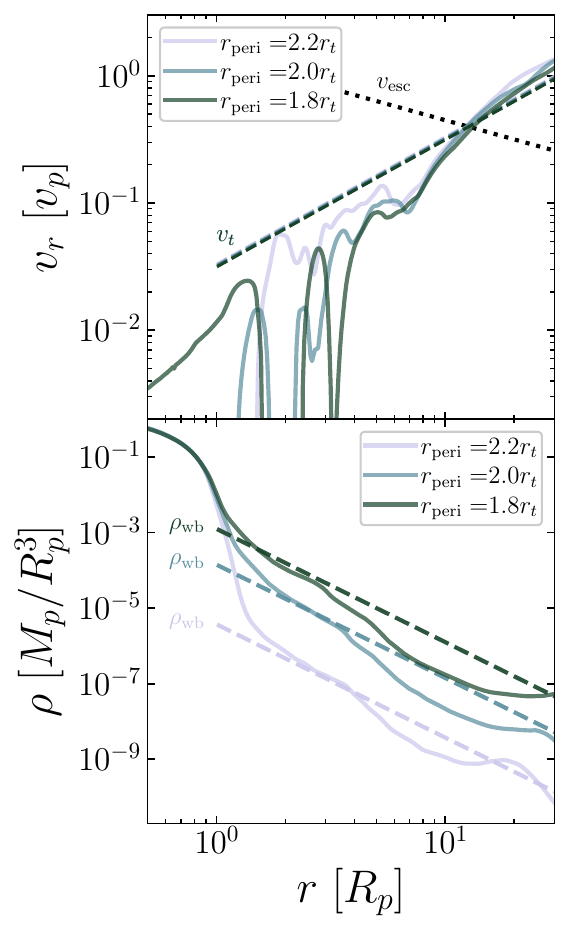}
  \caption{Tidal forces excite a radial velocity, driving an outflow. Radial velocity (top) and density (bottom) from our simulations are displayed for the $r_{\rm peri}$ values indicated, relative to $v_t$ (eq.~\ref{eq:v_t}), $v_{\rm esc} = (2G M_p/r)^{1/2}$, and $\rho_{\rm wb}$ (eq.~\ref{eq:rho_wb}). We calculate radial profiles at $t - t_{\rm peri} = 31 t_p$, when the planet Roche radius $R_{\rm RL} = 6.9, 7.0, 7.1 R_p$ for orbits with pericenter distances $r_{\rm peri} = 1.8, 2.0, 2.2 r_t$.
  }
  \label{fig:wind_prof}
\end{figure}

We compare the simulated power spectrum of radial flows at the planet surface, to the linear theory expectation, assuming power comes solely from the tidal excitation of normal modes:
\begin{equation}
    \mathcal{P}(\ell) = \frac{1}{4\pi} \sum_{m=-\ell}^\ell |v_{\ell m}|^2,
    \label{eq:PowerSpec}
\end{equation}
where
\begin{equation}
    v_{\ell m} = \Re\left[\int\limits_0^{2\pi} \int\limits_0^{\pi} v_r(R_p, \theta, \varphi)  Y_{\ell m}^*(\theta, \varphi) \sin\theta \,\der\theta \, \der\varphi \right].
\end{equation}
is the surface projection of $v_r$ onto spherical harmonics, with $\Re$ denoting the real part. After passage through pericenter, linear theory predicts
\begin{equation}
    |v_{\ell m}|^2 = |\Delta q|^2 \omega^2 |\tilde \xi_r(R_p)|^2 = \frac{\Delta E}{ {\widetilde M}},
    \label{eq:vlm_linear}
\end{equation}
with $\Delta E$ and $\widetilde M$ denoting the mode energy change and normalized mode mass:
\begin{align}
    \Delta E &= |\Delta q|^2 \tilde \varepsilon,
    \label{eq:DE_lin} \\
    \widetilde M &= \tilde\varepsilon \Big/\big[{\omega^2 |\tilde \xi_r(R_p)|^2}\big].
    \label{eq:tilde_M}
\end{align}
The dominant $\ell=2$ radial flow is well predicted by equation~\eqref{eq:vlm_linear} (Fig.~\ref{fig:power_spectrum} left panel). Before periastron ($t<t_{\rm peri}$), 
the power spectrum amplitudes grow as the planet nears the host. A precipitous drop in power is seen at large $\ell$, with numerical noise causing our simulation power to exceed the linear theory expectation.
After periastron ($t>t_{\rm peri}$), predicted $\mathcal{P}(2)$ and $\mathcal{P}(3)$ values are comparable to those from simulations. Shocks cause $\ell=2$ power to bleed into higher $\ell$. Although we find a similar Kolmogorov-like fall-off $\mathcal{P} \propto \ell^{-5/3}$ as \cite{MacLeod+2022}, higher power resides in even $\ell$ values, because the large $M_\star/M_p$ ratio considered here creates a strong symmetry to the tidal flow.

Shocks mediate the power transfer to higher angular degrees, stealing energy from the dominant $\ell = 2$ tidal oscillation (Fig.~\ref{fig:power_spectrum} right panel). The pericenter passage excites the $\ell=2$ fundamental mode to large amplitudes soon after $t_{\rm peri}$. Energy dissipation by shocks cause oscillation amplitudes to diminish with time, with deeper pericenter passages causing stronger shocks and more rapid amplitude decay. Past $10 t_{\rm peri}$, we find the decay rate to be approximately exponential, and fit
\begin{equation}
    \mathcal{P}(2) \propto \e^{-(t-t_{\rm peri})\gamma }
    \label{eq:P2_fit}
\end{equation}
to calculate the $\gamma$ values implied by our simulations. As expected, smaller $r_{\rm peri}$ have larger $\gamma$ (Fig.~\ref{fig:analytic_vs_sim_scaling} bottom panel). 
Damping rates exceed our grid's numerical noise and dissipation above $\gamma \gtrsim 10^{-3} t_p^{-1}$, constrained by simulating weak encounters without shocks in Appendix~\ref{sec:wide_peri}. We caution that the inferred $\gamma$ for $r_{\rm peri} = 2.2 r_t$ lies near the numerical dissipation limit. Simulations with $r_{\rm peri} \ge 2.2 r_t$ display oscillations in their $\ell=2$ power, at a frequency corresponding to the $\ell=2$ $f$-mode, due to the imperfect projection of the dominant tidal distortion onto $Y_{22}$.

Before and after periastron passage, the excited $f$-mode should change the total energy of the planet --- kinetic, gravitational, and internal --- by $\Delta E$. We compare the time average of the simulated total energy before time $t<5 t_p$, to that after $t - t_{\rm peri} > 80 t_p$, to the energy change predicted by equation~\eqref{eq:DE_lin} (Fig.~\ref{fig:analytic_vs_sim_scaling} top panel). The simulations agree with analytic estimates within a factor of $\sim$3.

In addition the excitation and deposition of energy, normal modes can also estimate the mass in the shocked atmosphere, and energy dissipation rate caused by wave breaking. Because acoustic waves break when their velocities are supersonic, the total mass contained in the shocked envelope should correspond to where the radial velocity of the dominant $f$-mode,
\begin{equation}
    v_r = \Im \left(2 \, \Delta q \, \omega \, \tilde \xi_r \, Y_{\ell m} \right),
    \label{eq:vr_lin}
\end{equation}
exceeds the local sound-speed $c_{\rm s}$, with $\Im$ denoting the imaginary part. Because the tidal response is dominated by $\ell, m = 2,2$ tidal component, this begins to occur above the radius $R_{\rm wb}$, where
\begin{equation}
    \left( \sqrt{ \frac{15}{8\pi} } \Delta q \, \omega \, \tilde \xi_r \right)_{R_{\rm wb}} = c_{\rm s},
    \label{eq:R_wb}
\end{equation}
evaluating $Y_{22}$ in equation~\eqref{eq:vr_lin} at the equator.  
Once eq.~\eqref{eq:R_wb} is satisfied, shocks sweep equatorially, heating material above $r>R_{\rm wb}$ (cf. Fig.~\ref{fig:peri_pass_profiles}, second row).
Comparing the mass enclosed in the region where $r$ exceeds $R_{\rm wb}$,
\begin{equation}
    M_{\rm wb} = \int_{R_{\rm wb}}^{R_p} 4\pi \rho r^2 \der r,
    \label{eq:M_wb}
\end{equation}
we find that equation~\eqref{eq:M_wb} agrees with the magnitude and scaling of the mass contained within the high-entropy region ($\ln[P/\rho^\Gamma]>0$) within a factor of $\sim$2 (Fig.~\ref{fig:analytic_vs_sim_scaling} middle panel).  
Because our grid has a minimum sound-speed $c_{\rm s} \approx 0.06 v_p$ near $R_p$, we expect large $r_{\rm peri}$ passages to excite surface gravity waves that do not break, confirmed in Appendix~\ref{sec:wide_peri}.

Energy dissipates when surface gravity waves steepen and shock. Taking the ratio of the energy dissipated from waves sweeping through the atmosphere, to the energy within the dominant $\ell=m=2$ oscillation mode \citep{MacLeod+2022}, wave breaking should damp energy at a rate
\begin{equation}
    \gamma_{\rm wb} \approx \epsilon_{\rm wb} \frac{\omega}{4\pi} \frac{M_{\rm wb}}{\widetilde M}.
    \label{eq:gamma_wb}
\end{equation}
Here, we neglect how breaking spins up the atmosphere, and throw in an order-unity fudge-factor $\epsilon_{\rm wb} = 0.3$ so that our order-of-magnitude scaling better matches simulations. The scaling predicted by equation~\eqref{eq:gamma_wb} agrees within a factor of a few with that seen in the power spectra decay rates  (Fig.~\ref{fig:analytic_vs_sim_scaling} bottom panel).

After waves break on the planet surface, we expect mass to escape the planet, entrained in a wind \citep[e.g.][]{MurrayClay+2009, OwenJackson2012}. Far from the planet, tidal forces dominate momentum balance, or
\begin{equation}
    v_r \frac{\partial v_r}{\partial r} \approx \frac{3 G M_\star r}{r_\star^3},
\end{equation}
causing the radial velocity to asymptote to 
\begin{equation}
    v_t = \left( \frac{3 G M_\star r^2}{r_\star^3} \right)^{1/2}.
    \label{eq:v_t}
\end{equation}
Our simulations find that $v_r$ nears $v_t$ past several $R_p$ (Fig.~\ref{fig:wind_prof} top panel). After some time, the mass loss rate should settle to a value that is roughly constant with radius. When most deposited energy supports a marginally-bound envelope, the mass loss rate $\dot M \propto \dot E_{\rm wb}$ in the absence of radiative losses \citep[e.g.][]{Quataert+2016}. Because the envelope initial mass is $M_{\rm wb}$, and mode-energy declines as $\propto \e^{-\gamma_{\rm wb}(t-t_{\rm peri})}$ (Fig.~\ref{fig:power_spectrum}), then the shock-powered mass loss rate should fall with time like
\begin{equation}
    \dot M \approx \gamma_{\rm wb} M_{\rm wb} \e^{-\gamma_{\rm wb}(t-t_{\rm wb})}.
\end{equation}
When mass loss is constant with distance, the continuity equation then gives the density:
\begin{equation}
    \rho_{\rm wb} \approx \frac{\gamma_{\rm wb} M_{\rm wb} }{4 \pi v_t r^2} \e^{-\gamma_{\rm wb}(t-t_{\rm wb})}.
    \label{eq:rho_wb}
\end{equation}
Estimate~\eqref{eq:rho_wb} predicts the magnitude of $\rho_{\rm wb}$, and its decline with $r$ and $t$, within factors of several (Fig.~\ref{fig:wind_prof} bottom panel). Disagreement is strongest where the outflow velocity exceeds the escape speed, and $\dot M$ is no longer spatially constant. Outflow densities for $r_{\rm peri} < 2.2 r_t$ are much larger than the artificial background, and are subject to little numerical contamination; outflow densities for $r_{\rm peri} \ge 2.2 r_t$ rival or are dwarfed by the background (App.~\ref{sec:wide_peri}).

\section{Planetary Structural and Orbital Evolution via Dynamical Tides}
\label{sec:Outcomes}

Section~\ref{sec:hydro_sim} demonstrates that close pericenter passages excite gas giant normal modes to such high amplitudes that they shock the surface, heating material to create an extended envelope. Wave breaking quickly diminishes fundamental-mode ($f$-mode) amplitudes, which act as a reservoir for orbital energy during high-eccentricity migration \citep[e.g.][]{VickLai2018, Wu2018, Vick+2019}. Here, we investigate the outcome of rapid $f$-mode damping over many proto-hot Jupiter orbits. Section~\ref{sec:MESA+GYRE} sets up our models for the structure and normal mode oscillations of gaseous planets. Section~\ref{sec:LongTerm} introduces our model for calculating the long-term evolution of planets whose $f$-modes break, and whether radiative diffusion can cool shocks and quench outflows. Section~\ref{sec:NepDesert} relates circularization via the stochastic growth and damping of $f$-mode energies, cooled by radiative diffusion or winds, to the demographic features of the Neptune Desert, Neptune Ridge, and hot-Jupiter pile-up. Section~\ref{sec:Photoevap} explores whether photoevaporation can remove residual envelopes retained following tidally-driven outflows.

\subsection{Planet Structure and Fundamental-Mode Models}
\label{sec:MESA+GYRE}

Because the tidal radius $r_t = R_p(M_\star/M_p)^{1/3} \propto \rho_p^{-1/3}$, the outcome of close periastron encounters depends on bulk density. For a solar composition envelope atop a compact core, degeneracy pressure causes the radius to become roughly fixed once gas dominates the mass of the planet ($R_p \sim R_{\rm Jup} \sim \text{constant}$). As mass increases, so does the bulk density, and planets must shorten their host separation to elicit the same tidal response. On the other hand, once a planet's mass becomes dominated by its core, 
planet sizes decrease and bulk densities climb as the gas mass falls. Because an envelope with just a few percent of the mass of the planet can double the radius \citep[e.g.][]{OwenWu2017}, even a slight amount of gas drastically alters the size of a Neptune.

Here, we discuss our structure models used to ascertain how planets respond to impulsive tidal encounters. We use \texttt{MESA} to model the structure and evolution of our planets. We initialize constant entropy planets with $13 M_{\rm Jup}$ or $1 M_{\rm Jup}$ masses, placing a $M_{\rm core} = 10 M_\oplus$ inert core at the center, choosing a $10 \, {\rm g}/{\rm cm}^3$ core density based off expectations from an Earth-composition Silicon-Iron equation of state \citep[e.g.][]{Fortney+2007}. We use \texttt{relax\_mass\_scale} to gradually change the planet mass while keeping its composition as a function of $M(<r)/M_p$ constant. Our models of super-Jupiter planets strip an initial $13 M_{\rm Jup}$ body to masses that span a 20 point logarithmic grid between $13 M_{\rm Jup}$ and $1.1 M_{\rm Jup}$, while our models of Jovian and sub-Saturn planets strip a $1 M_{\rm Jup}$ body to envelope masses $M_{\rm env}$ that span a 100 point logarithmic grid between $1 M_{\rm Jup}$ and $0.01 M_\oplus$. We note that we clip the lowest 11 points from the sub-Saturn grid due to \texttt{MESA} timestep issues. Each planet is then irradiated with a dayside flux of $2.5 \times 10^8 {\rm erg}/{\rm cm}^2/{\rm s}$, corresponding to a 2.6 day orbital period around a solar-luminosity star\footnote{
The dayside flux averaged over the orbit of an eccentric gas giant is typically smaller than $2.5 \times 10^8 {\rm erg}/{\rm cm}^2/{\rm s}$: e.g. $\{a, e\}=\{0.5 \ {\rm AU}, 0.98\}$ around a solar host gives $2.7 \times 10^7 {\rm erg}/{\rm cm}^2/{\rm s}$, using eq.~1 of \cite{Bolmont+2016}. Our adoption of $2.5 \times 10^8 {\rm erg}/{\rm cm}^2/{\rm s}$ is closer to the dayside flux at pericenter, when the $f$-mode is excited and dissipates most strongly. Because the photosphere sound-speed $c_{\rm s} \propto T_{\rm eq}^{1/2} \propto D^{1/4}$, a high $e=0.98$ will cause $c_s$ to vary by at most factor of $\sim 3$.
}, to a column depth of $300 \ {\rm cm}^2/{\rm g}$. We integrate the planet structure equations out to the photosphere ($\tau = 2/3$), which we take to be the planet radius $R_p$. We then evolve each model for $10^9$ years. At $10^9$ years, the many Kelvin-Helmholtz times that have elapsed imply the planet structure depends negligibly on the initial entropy \citep[e.g.][]{Marley+2007, OwenWu2016}. The energy required to heat our solid core is accounted by adding the extra luminosity at the envelope base
\begin{equation}
    L_{\rm ext} = c_V M_{\rm core} \frac{\der T_{\rm core}}{\der t},
\end{equation}
where the rock/iron core specific heat $c_V = 10^7 \ {\rm erg}/{\rm g}/{\rm K}$ \citep[e.g.][]{Valencia+2010}, and the core temperature is the same as the envelope base $T_{\rm core} = T(R_{\rm core})$. We neglect radiogenic heating for simplicity \citep[e.g.][]{ChenRogers2016, HallattMillholland2025}.

\begin{figure}
  \centering
  \includegraphics[width=\columnwidth]{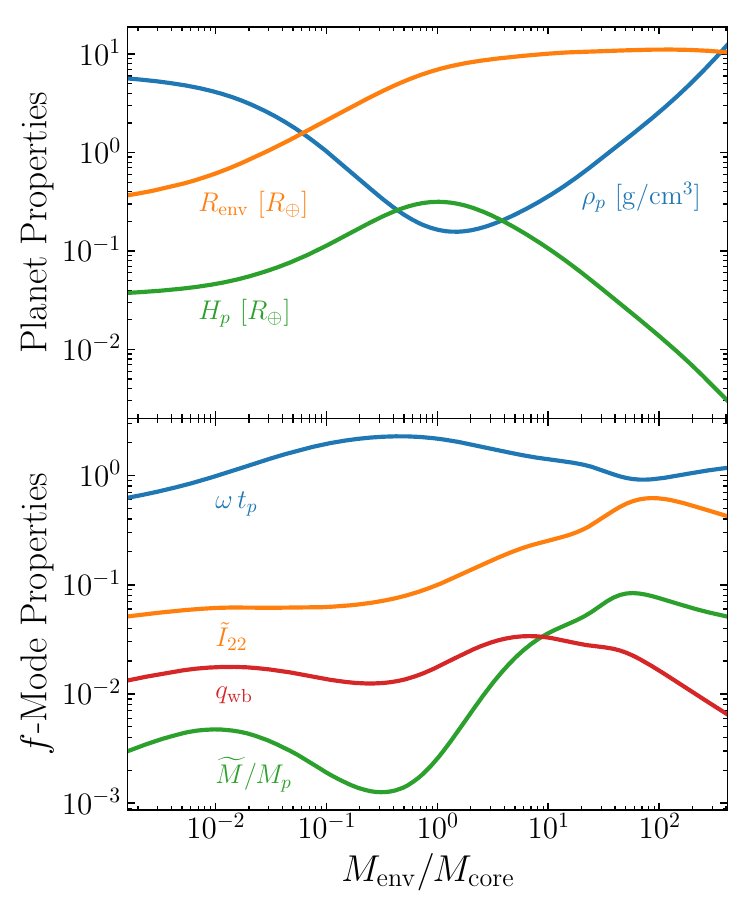}
  \caption{Planet (top) and fundamental-mode (bottom) properties from \texttt{MESA} and \texttt{GYRE}, plotted against the planet mass, displayed by the envelope to core mass ratio $M_{\rm env}/M_{\rm core}$ for a $M_{\rm core} = 10 M_\oplus$ core. Evaluated at an age of 1 Gyr, all quantities are constant with time when used in the iterative map detailed in Section~\ref{sec:LongTerm}. Envelope radius $R_{\rm env} = R_p-R_{\rm core}$, surface density scale-height $H_p = (\rho |\der \rho/\der r|^{-1})_{R_p}$, and bulk density $\rho_p = 3 M_p/(4\pi R_p^3)$ displayed in units indicated; mode-to-dynamical frequency ratio $\omega \, t_p$, overlap coefficient $\tilde I_{22}$, mode-to-planet mass ratio $\widetilde M/M_p$, and wave-breaking amplitude $q_{\rm wb} = (c_{\rm s}/v_r)_{R_p}$, are dimensionless.
  }
  \label{fig:mesa_gyre_plot}
\end{figure}

Key properties of our planet models are displayed in the top panel of Figure~\ref{fig:mesa_gyre_plot}. Envelope radii $R_{\rm env} = R_p - R_{\rm core}$ increase as the gas-to-core mass ratio $M_{\rm env}/M_{\rm core}$ increases, then asymptote to $R_{\rm env} \sim 9 R_\oplus$ due to degeneracy pressure once $M_{\rm env}/M_{\rm core} \gtrsim 1$. This causes the $\sim$0.1 ${\rm g}/{\rm cm}^3$ minimum in the bulk density when the envelope and core masses are comparable ($M_{\rm env}/M_{\rm core} \sim 1$), which rises to $\sim$10 ${\rm g}/{\rm cm}^3$ when $M_{\rm env}$ is much smaller, or larger, than $M_{\rm core}$. Similarly, the density scale-height at the surface $H_p$ has a maximum when $M_{\rm env}/M_{\rm core} \sim 1$.

We neglect the small variation in planetary density caused by the short burst of irradiation received at pericenter. A day-side flux $F_{\rm irr} = 2.5 \times 10^8 \, {\rm erg}/{\rm cm}^2/{\rm s}$ increases sub-Saturn 
temperatures at shallow depths, otherwise temperatures drop below $100 \, {\rm K}$ and the \texttt{MESA} H/He equation of state breaks down. Day-side fluxes below $\lesssim 10^9 \, {\rm erg}/{\rm cm}^2/{\rm s}$ that penetrate to column depths $\lesssim 300 \, {\rm cm}^2/{\rm g}$ have little impact on gaseous planet radii \citep[e.g.][]{LopezFortney2014, KomacekYoudin2017}.

The stellar oscillation code \texttt{GYRE} \citep{TownsendTeitler2013} is used to linearly perturb the equations of hydrostatic equilibrium for each planet model, and calculate the properties of $f$-modes for different masses. We focus on the dominant $f$-mode with a radial node number $n_{\rm pg} = 0$ and angular degree $\ell=2$, and neglect Coriolis forces. We assume modes are adiabatic, since thermal times are longer than $f$-mode periods within giant planets \citep[e.g.][]{Zanazzi+2025, Fuller+2026}. We set the radial displacement to zero at the core boundary, and the pressure Lagrangian displacement to zero at the surface. Equations are integrated using the \texttt{MAGNUS\_GL4} difference scheme. 

The $f$-mode properties with the envelope-to-core mass-ratio are displayed in the bottom panel of Figure~\ref{fig:mesa_gyre_plot}. The $f$-mode frequency is comparable to the dynamical frequency over all planet models considered ($\omega t_p \sim 1$). Two other quantities --- the overlap integral $\tilde I_{22}$ (eq.~\ref{eq:I_lm}), and mode-to-planet mass-ratio $\widetilde M/M_p$ (eq.~\ref{eq:tilde_M}) --- can vary non-trivially with mass when $M_{\rm env} \gtrsim 10 M_{\rm core}$. This is driven by ionization of hydrogen when the planet interior exceeds $\gtrsim 10^4 \, {\rm K}$, which causes abrupt variations in the adiabatic index and sound-speed $c_{\rm s}$, affecting the $f$-mode radial wavenumber $k_r \sim \omega/c_{\rm s}$. 

Figure~\ref{fig:mesa_gyre_plot} also displays the critical mode amplitude
\begin{equation}
    q_{\rm wb} = \sqrt{ \frac{8\pi}{15} } \left( \frac{c_{\rm s}}{\omega \tilde \xi_r} \right)_{R_p},
\end{equation}
obtained by setting $R_{\rm wb} = R_p$ in equation~\eqref{eq:R_wb}, above which $f$-modes dissipate energy through surface shocks. The amplitude hovers at a value slightly above $1\%$ for most of our models, until the planet mass $M_p \gtrsim 500 M_\oplus$, where $q_{\rm wb}$ drops because of the lower surface sound-speeds on our super-Jupiters. This corresponds to a mode energy $E_{\rm wb} = |q_{\rm wb}|^2 E_p  \lesssim 10^{-3} E_p$, much less than the $E_{\rm crit} \sim 10^{-2} - 10^0 E_p$ values used in previous studies of $f$-mode circularization, where non-linear dissipation was assumed to rapidly damp mode energies to zero \citep[e.g.][]{VickLai2018, Wu2018}. As we will see, these comparatively small $E_{\rm wb}$ values from wave breaking leash the $f$-mode random walk, causing the mode energy to diminish long before it nears the binding energy. Circularization still proceeds with small $E_{\rm wb}$, but $a$ and $e$ undergo less dramatic changes.

\subsection{Dynamical Tides, Dissipation, and Mass Loss}
\label{sec:LongTerm}

\begin{figure*}
  \centering
  \includegraphics[width=\linewidth]{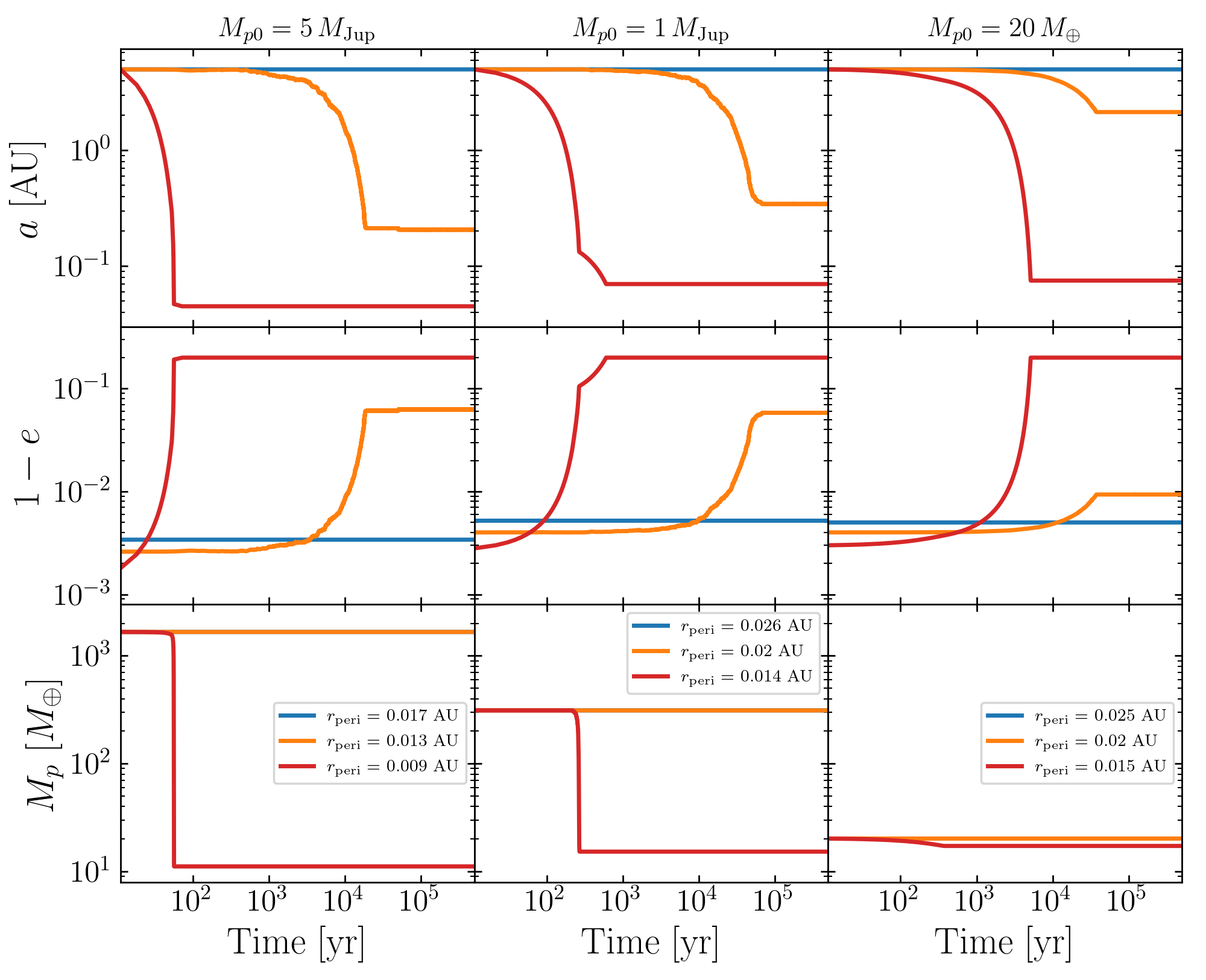}
  \caption{Once long-period planets are excited to high eccentricities, dynamical tides circularize orbits, with shocks driving outcomes that depend sharply on the pericenter distance $r_{\rm peri}$. Semi-major axis $a$ (top), eccentricity $e$ (middle), and mass $M_p$ (bottom) evolution with time are displayed in different rows, for planets whose orbits have $r_{\rm peri}$ values that place them in the stagnant (blue), diffusive (orange), and outflow (red) regimes. Columns display calculations with indicated initial planet masses $M_{p0}$, all with the same initial semi-major axes $a_0 = 5 \ {\rm AU}$. We freeze orbital elements and mode energies once $e<0.8$.
  }
  \label{fig:orb_ev}
\end{figure*}

\begin{figure}
  \centering
  \includegraphics[width=\columnwidth]{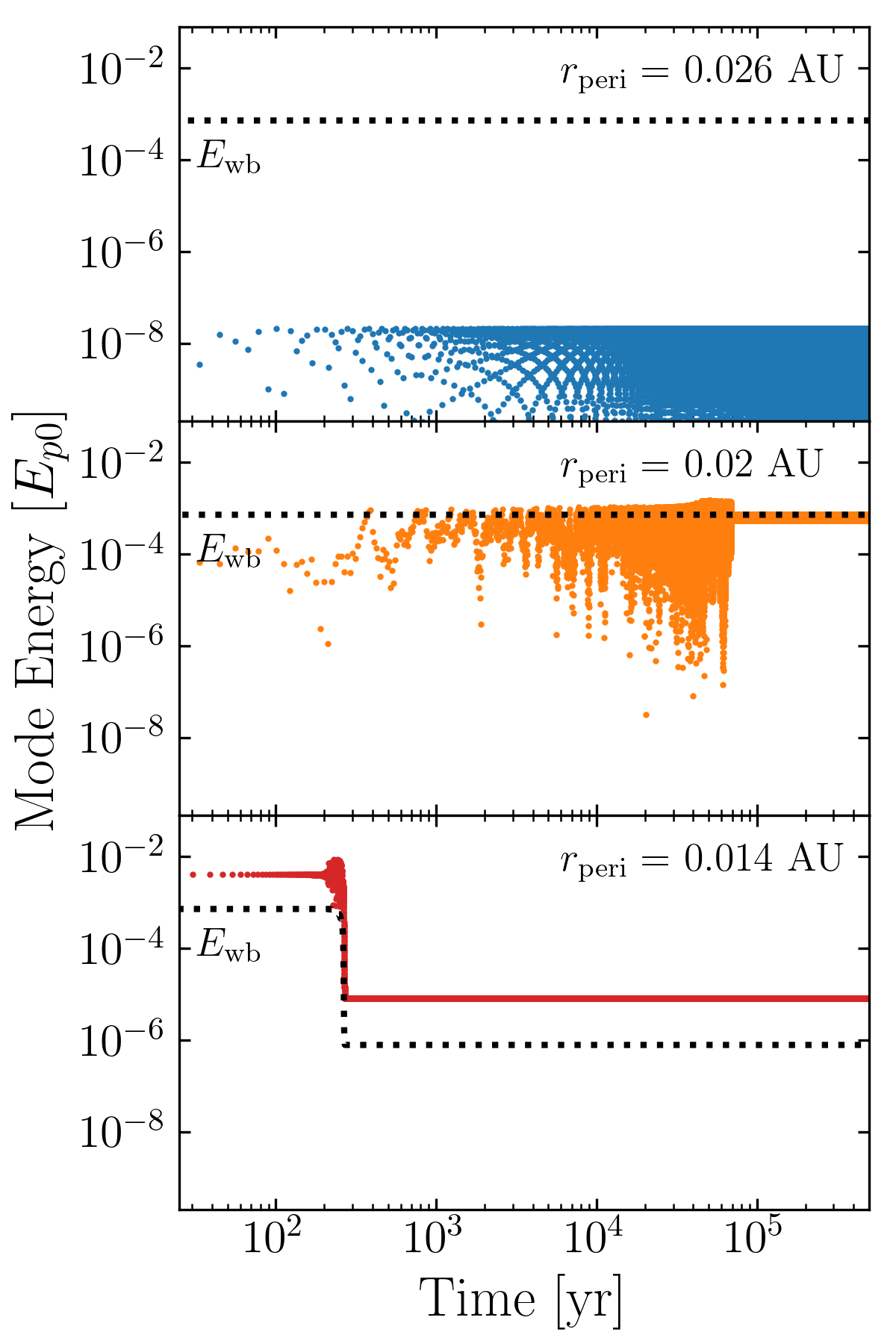}
  \caption{Mode energy evolution for a Jovian-mass body, with $r_{\rm peri}$ values that place the planet in three different dynamical tide regimes. Large $r_{\rm peri}$ have stagnant mode energies that don't grow (``stagnant regime,'' top), closer $r_{\rm peri}$ stochastically grow mode energies that cause shallow shocks that diffusively cool (``diffusive regime,'' middle), while the shortest $r_{\rm peri}$ excite modes that shock deep and drive outflows every orbit (``outflow regime,'' bottom). Modes dissipate when their energies exceed the wave-breaking energy $E_{\rm wb} = |q_{\rm wb}|^2 E_p$, displayed by dotted lines. Energies displayed in units of the initial binding energy $E_{p0} = G M_{p0}^2/R_{p0}$.
  }
  \label{fig:energy_ev}
\end{figure}

The circularization of hot Jupiters by dynamical tides can be modeled with an iterative map \citep{Mardling1995b}. When a gas giant on an eccentric orbit passes through pericenter, the $f$-mode gains energy from the encounter. For highly eccentric orbits ($r_{\rm peri} \ll a$), the $f$-mode excitation is modeled as impulsive kick, at times $t_k$ with iterative indices $k$ corresponding to periastron passages. Mode amplitudes before $q_{k}$ and after $q_k'$ closest approach are related by
\begin{equation}
    q_k' = q_{k} + \Delta q_{k},
    \label{eq:qk_map}
\end{equation}
where $\Delta q_{k}$ for $\ell = m = 2$ is given by equation~\eqref{eq:Dq}, using the \cite{Lai1997} approximation for $K_{22}$:
\begin{equation}
    K_{22} \approx \frac{2 z^{3/2} \e^{-2z/3}}{\Omega_{\rm peri} t_p \sqrt{15}} \left( 1 - \frac{1}{4} \sqrt{ \frac{\pi}{z} } \right),
    \label{eq:K_22_approx}
\end{equation}
with $z = \sqrt{2}\omega/\Omega_{\rm peri}$, and $\Omega_{\rm peri} = (G M_\star/r_{\rm peri}^3)^{1/2}$. The sudden kick at $t_k$ causes the mode energy $E_k = |q_k|^2 E_p$ to change by
\begin{equation}
    \Delta E_k = \left(|q_k'|^2 - |q_{k}|^2 \right)E_p.
\end{equation}
Energy is transferred between the oscillation and the orbit. If the orbital energy immediately before $t_k$ is $E_{{\rm orb},k-1}$, the new orbital energy is given by
\begin{equation}
    E_{{\rm orb},k} = E_{{\rm orb},k-1} - \Delta E_k.
\end{equation}
If the planet mass after $t_k$ is $M_{p,k}$, its semi-major axis is
\begin{equation}
    a_k = -\frac{G M_\star M_{p,k}}{2 E_{{\rm orb},k}},
\end{equation}
with the orbital period $P_{{\rm orb},k} = 2\pi [a_k^3/(G M_\star)]^{1/2}$. However, $r_{\rm peri}$ remains constant during these instantaneous pericenter kicks, so
\begin{equation}
    e_k = 1 - r_{\rm peri}/a_k.
\end{equation}
Once kicked, the free oscillation rings and damps at new amplitude $\propto \e^{-\im \omega t - \gamma t}$, until right before its subsequent periastron kick at $t_{k+1} = t_k + P_{{\rm orb},k}$, where
\begin{equation}
    q_{k+1} = q_k' \e ^{ -\im \omega_k P_{{\rm orb},k} - \gamma_k P_{{\rm orb},k} },
    \label{eq:q_k'}
\end{equation}
with $\omega_k$ and $\gamma_k$ being the mode frequency and damping rate after $t_k$. Following $t_{k+1}$, the mode amplitude is given by $q_{k+1}' = q_{k+1} + \Delta q_{k+1}$, and the map repeats. Mode amplitudes are initialized by $\{q_0, t_0\} = \{0, P_{\rm orb,0}\}$. Because approximation~\eqref{eq:K_22_approx} gives $K_{22}$ in terms of $\omega_k$ and $r_{\rm peri}$, while the other terms in $\Delta q_k$ depend only on the planet and mode properties (eq.~\ref{eq:Dq}), $\Delta q_k$ varies with $k$ only when the planet changes mass.

Even when individual kicks change mode energies by a small amount $|\Delta E_k| \ll E_{\rm wb}$, if the mode phase between $t_k$ and $t_{k+1}$ is uncorrelated, mode energies $E_k$ can stochastically grow through a random walk to exceed $E_{\rm wb}$ after many orbits. Inspecting equation~\eqref{eq:q_k'}, this occurs when
\begin{align}
    &\left| \omega_k P_{{\rm orb},k+1} - \omega_{k} P_{{\rm orb},k} \right| \approx \omega_k \left| \Delta P_{{\rm orb},k} \right|
    \nonumber \\
    &\approx \frac{3}{2} \omega_k P_{{\rm orb},k} \left| \Delta q_k \right|^2 \left| \frac{E_{p,k}}{E_{{\rm orb},k}} \right| \gtrsim 1,
    \label{eq:q_grow}
\end{align}
where $E_{p,k} = G M_{p,k}^2/R_{p,k}$ is the planet binding energy and $R_{p,k}$ the planet radius at iteration $k$. Left unhindered, eccentric hot Jupiters that satisfy equation~\eqref{eq:q_grow} can grow their $f$-modes to an appreciable fraction of their binding energy \citep[e.g.][]{VickLai2018, Wu2018}. 

However, once $q_k$ exceeds $q_{\rm wb}$, $f$-modes shock the atmosphere and dissipate energy at a rate
\begin{equation}
    \dot E_{\rm wb} \approx \gamma_{\rm wb} |q_k|^2 E_p.
\end{equation}
The response of the atmosphere depends on the relative value of $\dot E_{\rm wb}$ to the Eddington luminosity
\begin{equation}
    L_{\rm Edd} = \frac{4\pi G M_{p,k} c}{\kappa}.
\end{equation}
When $\dot E_{\rm wb} < L_{\rm Edd}$, the photon diffusion time is short, and the atmosphere radiatively cools without loosing mass between $t_k$ and $t_{k+1}$.\footnote{The condition $\dot E_{\rm wb} < L_{\rm Edd}$ is equivalent to the photon diffusion time $t_{\rm diff} \sim \tau R_p/c$ being shorter than the outflow advection time $t_{\rm adv} \sim R_p/v_r$ \citep{Quataert+2016}.} However, if $\dot E_{\rm wb} > L_{\rm Edd}$, wave breaking drives a hydrodynamical wind \citep[e.g.][]{Quataert+2016}. Because shocks heat the atmosphere to temperatures that exceed $\gtrsim 10^4$ K, electron scattering dominates the opacity ($\kappa \approx 0.34 \ {\rm cm}^2/{\rm g}$). 

Our iterative map decreases the planet mass when shocks unbind material above $R_{\rm wb}$. We substitute $q_k$ for $\Delta q$ in the definition of $R_{\rm wb}$ (eq.~\ref{eq:R_wb}), and include the the mass above the photosphere when calculating $M_{\rm wb}$ in $\gamma_{\rm wb}$ (eq.~\ref{eq:gamma_wb}):
\begin{equation}
    M_{\rm wb} = \int_{R_{\rm wb}}^{R_p} 4\pi \rho r^2 \der r + 4\pi R_p^2 \rho(R_p) H_p.
\end{equation}
Here, $H_p = (\rho |\der \rho/\der r|^{-1})_{R_p}$ is the density scale-height at the surface. Because the loss of mass $|\Delta M_p| \ll M_p$ changes the potential energy of the planet by
\begin{equation}
    \Delta E_p = -\frac{G M_p}{R_p} \Delta M_p,
\end{equation}
energy conservation limits $|\Delta M_p| \le R_p \Delta E /(G M_p)$. We therefore change the mass at orbit $k$ using
\begin{equation}
    \Delta M_{p,k} = M_{p,k} - M_{p,k-1} = -\min \left( M_{\rm wb}, \frac{R_{p,k} \Delta E_{k}}{G M_{p,k}} \right),
\end{equation}
unless $\Delta E_k<0$, where we force $\Delta M_{p,k} = 0$. The mode damping rate $\gamma_k = 0$ when $q_k < q_{\rm wb}$, and $\gamma_k = \gamma_{\rm wb}/2$ when $q_k \ge q_{\rm wb}$ (the $\frac{1}{2}$ factor arises because $\gamma_{\rm wb}$ is the energy damping rate $\dot E \propto \e^{-\gamma_{\rm wb}t} \propto |q|^2$). When tracking the orbit and mass evolution, we calculate $\Delta q_k$ by picking a stellar and normal mode model whose mass lies closest to $M_p$ every iteration (Fig.~\ref{fig:mesa_gyre_plot}). We also halt maps when the eccentricity falls below $e<0.8$, where the assumption of constant $r_{\rm peri}$ breaks down (see eqs.~\ref{eq:DJorb_over_Jorb}-\ref{eq:DJ_over_Jorb} below).

Circularization by dynamical tides depends strongly on the pericenter separation $r_{\rm peri}$ (Fig.~\ref{fig:orb_ev}). At large $r_{\rm peri}$, the relative mode phase shifts negligibly between orbits ($\omega \Delta P_{\rm orb}<1$, eq.~\ref{eq:q_grow}), and mode amplitudes do not stochastically grow (Fig.~\ref{fig:energy_ev}, ``stagnant regime''). Energies $E_k$ remain well-below the wave-breaking energy $E_{\rm wb} = |q_{\rm wb}|^2 E_p$ (Fig.~\ref{fig:energy_ev} top panel), and we recover the quasi-periodic variations in $E_k$ of \cite{Mardling1995a, IvanovPapaloizou2004, VickLai2018, Wu2018}. Little energy is deposited in oscillations, and the orbital evolution stagnates (Fig.~\ref{fig:orb_ev} blue lines). At intermediate $r_{\rm peri}$, the phase shift between orbits causes $q_k$ to grow ($\omega \Delta P_{\rm orb}>1$), but shallow wave breaking deposits heat in an atmosphere that cools through radiative diffusion (Fig.~\ref{fig:energy_ev} middle panel, ``diffusive regime''). Mode energies $E_k$ quickly damp once they exceed $E_{\rm wb}$, but most gas is retained from one periastron passage to the next. Normal modes sap orbital energy until successive kicks no longer occur at different phases ($\omega \Delta P_{\rm orb} \lesssim 1$), halting the diffusive growth of $q_k$, and freezing the orbit (Fig.~\ref{fig:orb_ev} orange lines). 


For small $r_{\rm peri}$, shocks are sufficiently strong and deep to drive outflows. Every periastron passage excites $E_k$ to an amplitude that shocks atmospheres at super-Eddington rates ($\dot E_{\rm wb} > L_{\rm Edd}$), reliably unbinding mass every orbit (Fig.~\ref{fig:energy_ev} bottom panel, ``outflow regime''). Analogous to Roche-lobe overflow \citep[e.g.][]{HjellmingWebbink1987}, when the variation of the wave-breaking location with mass, or
\begin{equation}
    \zeta_{\rm wb} = \frac{\der \ln R_{\rm wb}}{\der \ln M_p},
    \label{eq:zeta_wb}
\end{equation}
is larger than that of planet radius
\begin{equation}
    \zeta_p = \frac{\der \ln R_p}{\der \ln M_p},
    \label{eq:zeta_p}
\end{equation}
mass loss occurs rapidly ($\zeta_{\rm wb} > \zeta_p$, Fig.~\ref{fig:orb_ev}, red lines). Outflows quickly unbind gas giant envelopes because degeneracy pressure causes $\zeta_p \approx 0$. Eventually $R_{\rm wb}$ is so deep that the fractional change in gas mass per orbit exceeds $|\Delta M_{\rm env}|/M_{\rm env} \gtrsim 10^{-2} - 10^{-1}$, with most mass lost in $\sim 10-100$ orbits. Outflows abruptly halt once planets become sufficiently core-dominated ($M_{\rm env}/M_{\rm core} \lesssim 10^{-1} - 1$), and gas pressure supports envelopes causing $\zeta_p \gg 1$. Wave breaking on tenuous envelopes continues to dissipate energy and circularize orbits, but no longer unbinds mass past $\gtrsim 10^2$~years (Fig.~\ref{fig:orb_ev}, red lines).

\begin{figure}
  \centering
  \includegraphics[width=\columnwidth]{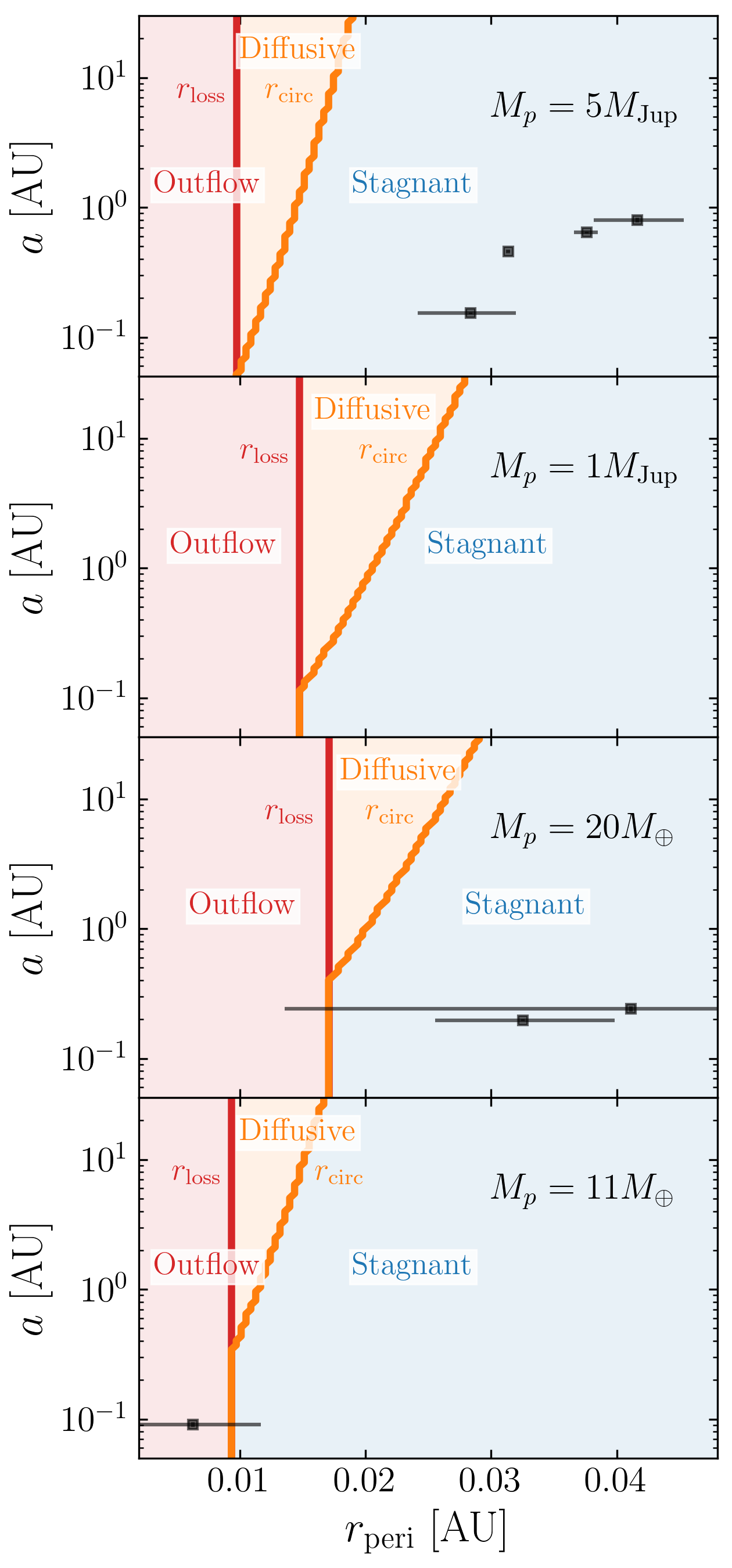}
  \caption{Whether tidal oscillations lie in the stagnant (blue), diffusive (orange), or outflow (red) regimes depends on the semi-major axis $a$ and pericenter distance $r_{\rm peri}$ of the planet. We mark the largest pericenter separations where dynamical tides can circularize orbits ($r_{\rm circ}$) and drive outflows ($r_{\rm loss}$). Different panels display results for model planet masses, with black dots denoting eccentric ($e>0.8$) gas giants (masses $10 M_\oplus$ to $13 \Mjup$), which we categorize as super-Jupiters ($\sim 5 M_{\rm Jup}$; HD 80606b \citealt{Pearson+2022}, HIP 66074b \citealt{Sozzetti+2023}, TIC-241249530b \citealt{Gupta+2024}, TOI-3362b \citealt{Dong+2021}), sub-Saturns ($\sim 20 M_\oplus$; HD-144899b \citealt{Feng+2022},  Kepler-1656b \citealt{Angelo+2022}), and sub-Neptunes ($\sim 11 M_\oplus$; GJ-3222b \citealt{Feng+2022}).
  }
  \label{fig:radii_diag}
\end{figure}

The pericenter separations where mode energies remain stagnant, diffusively grow, and drive outflows, depends strongly on bulk density. We define $r_{\rm circ}$ as the $r_{\rm peri}$ when $\omega \Delta P_{\rm orb} = 1$, and $r_{\rm loss}$ as the $r_{\rm peri}$ when $\dot E_{\rm wb} = L_{\rm Edd}$ --- when $r_{\rm peri}$ is smaller than $r_{\rm circ}$, circularization can commence through the stochastic growth of $f$-modes, and when $r_{\rm peri}$ is less than $r_{\rm loss}$, shocks drive mass loss.\footnote{
If $\gamma_{\rm wb}^{-1}$ is longer than $P_{\rm orb}$, it is in-principle possible for mode amplitudes to grow from the diffusive regime into the outflow regime. In practice however, we find that $\gamma_{\rm wb}^{-1} \gg P_{\rm orb}$ when $\dot E_{\rm wb}$ is comparable to $L_{\rm Edd}$.}
Because both $r_{\rm circ}$ and $r_{\rm loss}$ scale with the tidal radius $r_t \propto \rho_p^{-1/3}$, denser planets generally need closer approaches to circularize and remove their massive envelopes (Fig.~\ref{fig:radii_diag}). Bulk densities are smallest for when the envelope mass is comparable to that of the core ($M_p \sim 20 M_\oplus$), maximizing both $r_{\rm circ}$ and $r_{\rm loss}$. Planet densities increase as $M_p$ strays from $\sim20M_\oplus$ (Fig.~\ref{fig:mesa_gyre_plot}), causing both $r_{\rm circ}$ and $r_{\rm loss}$ to move closer to the host star. The circularization boundary $r_{\rm circ}$ scales with semi-major axis $a$, since longer orbital periods cause mode phase offsets to be larger between successive pericenter kicks. The mass loss boundary $r_{\rm loss}$ is insensitive to $a$, because we assume encounters are nearly parabolic, taking $K_{22}$ to depend only on $r_{\rm peri}$ and $\omega$ (eq.~\ref{eq:K_22_approx}).

The short timescale by which orbits circularize via dynamical tides imply that few systems should be seen in the diffusive or outflow regimes. Comparing our expectations to eccentric ($e > 0.8$) gaseous planets (masses $10 M_\oplus$ to $13 \Mjup$) listed on the NASA Exoplanet Archive, we confirm that highly-eccentric super-Jupiters like HD 80606b ($e=0.93$, \citealt{Pearson+2022}) HIP66074b ($e = 0.95$, \citealt{Sozzetti+2023, Fitzmaurice+2026}) and TIC-241249530b ($e = 0.94$, \citealt{Gupta+2024}) lie in the stagnant regime, whose pericenters are too large to stochastically grow $f$-mode energies \citep[see also][]{Liveoak+2026b}. 
However, we note that the $\sim 11 M_\oplus$ eccentric ($e = 0.93$) sub-Neptune GJ-3222b might currently be migrating by dynamical tides \citep{Feng+2022}.


Because $r_{\rm peri}$ remains constant throughout our iterative map, we violate momentum conservation \citep[e.g.][]{Mardling1995a}. However, as long as the angular momentum exchanged between the orbit and oscillation are small, our constant $r_{\rm peri}$ assumption is justified. Because constant $r_{\rm peri}$ implies that small changes in the semi-major axis $\Delta a$ and eccentricity $\Delta e$ are related by
\begin{equation}
    \frac{\Delta e}{1-e} = \frac{\Delta a}{a},
\end{equation}
our map changes the orbital angular momentum $J_{\rm orb} = [G M_\star (1+e) r_{\rm peri}]^{1/2}$ by
\begin{equation}
    \frac{\Delta J_{\rm orb}}{J_{\rm orb}} = - \frac{1-e}{2(1+e)} \frac{\Delta E_{\rm orb}}{E_{\rm orb}}.
    \label{eq:DJorb_over_Jorb}
\end{equation}
Similarly, because normal modes have angular momenta $J = \frac{m}{\omega} E$, $m=2$ maps change $J$ by
\begin{equation}
    \frac{\Delta J}{J_{\rm orb}} = - \frac{1-e}{\sqrt{1+e}} \frac{\Omega_{\rm peri}}{\omega} \frac{\Delta E}{E_{\rm orb}}.
    \label{eq:DJ_over_Jorb}
\end{equation}
As long as the orbit is highly eccentric ($|1-e| \ll 1$), changes in angular momentum are small relative to those in energy ($|\Delta J_{\rm orb}/J_{\rm orb}| \ll |\Delta E_{\rm orb}/E_{\rm orb}|$, $|\Delta J/J_{\rm orb}| \ll |\Delta E/E_{\rm orb}|$). Halting iterations when $e<0.8$ prevents the neglect of $\Delta J_{\rm orb}$ from being problematic.

\subsection{Hot Jupiter Pileup, Neptune Ridge, and Neptune Desert from Dynamical Tides}
\label{sec:NepDesert}

\begin{figure*}[htbp]
  \centering
  \includegraphics[width=\textwidth]{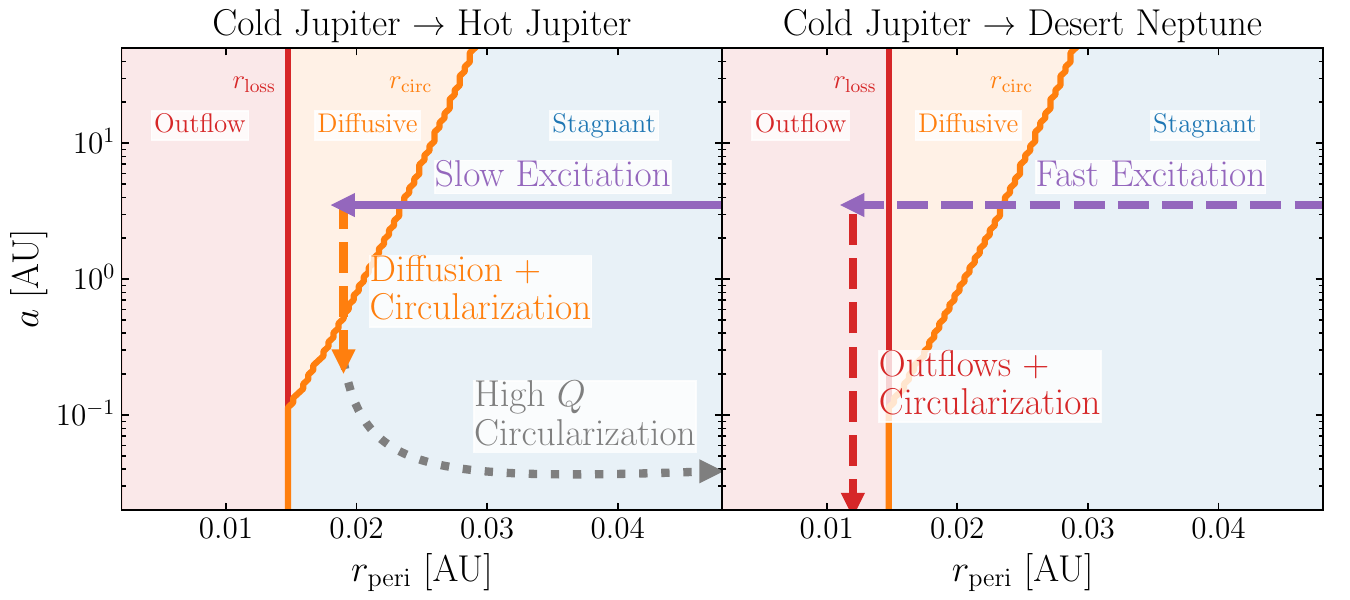}
  \caption{
  Whether gas giants become hot Jupiters or desert Neptunes following high-eccentricity migration depends on the timescale eccentricity is excited $t_{\rm excite}$, relative to the orbit decay timescale from dynamical tides $t_{\rm decay} \sim P_{\rm orb} |E_{\rm orb}/\Delta E|$.
  If $t_{\rm excite}$ is longer than $t_{\rm decay}$ (left panel), $r_{\rm peri}$ stops decreasing when $r_{\rm peri} < r_{\rm loss}$ (solid purple arrow). The $f$-mode shocks cool via radiative diffusion, Jovians don't loose mass, and hot-Jupiter orbits pile-up interior to $r_{\rm circ}$ after circularization by dynamical tides (dashed orange arrow). The final semi-major axis is attained after circularization via tidal dissipation with a higher $Q$ (dotted gray lines). If $t_{\rm excite}$ is shorter than $t_{\rm decay}$ (right panel), $r_{\rm peri}$ decreases until $r_{\rm peri} < r_{\rm loss}$ (dashed purple arrow). The $f$-mode shocks drive outflows, causing the gas giant to loose mass while it circularizes via dynamical tides (dashed red arrow). A remnant sub-Saturn core lies within the Neptune desert after circularization. We display $r_{\rm loss}$ and $r_{\rm circ}$ values for a Jovian planet with a solar host; Stagnant/Diffusive/Outflow regime labels and colors are identical to Figure~\ref{fig:radii_diag}.
  }
  \label{fig:HJ_vs_DN_formation}
\end{figure*}

\begin{figure*}[htbp]
  \centering
  \includegraphics[width=\textwidth]{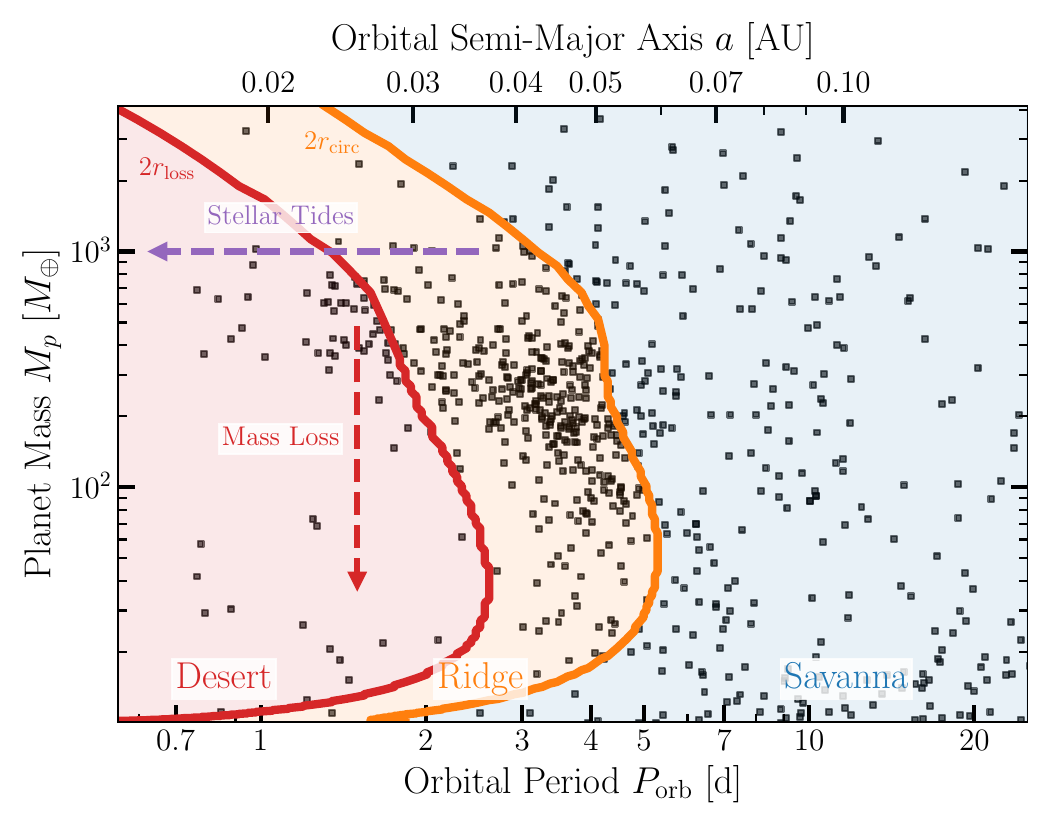 }
  \caption{
  Mass loss during high-eccentricity migration puts gas giant cores in the Neptune desert interior to $2 r_{\rm loss}$, while stochastically-excited $f$-modes can cluster planets at the hot-Jupiter pile-up and Neptune ridge between $2 r_{\rm loss}$ and $2 r_{\rm circ}$. Small periastron distances $r_{\rm peri}$ excite $f$-modes that circularize orbits and deposit heat through shocks, which cool through outflows when $r_{\rm peri} < r_{\rm loss}$ (red), or radiative diffusion when $r_{\rm loss} < r_{\rm peri} < r_{\rm circ}$ (orange). At large $r_{\rm peri} > r_{\rm circ}$ (blue), $f$-modes are not stochastically-excited, and their stagnant amplitudes don't damp eccentricities. Complete circularization parks planets at distances $2 r_{\rm peri}$. We calculate $r_{\rm circ}$ for an initial semi-major axes of 10~AU (see Fig.~\ref{fig:radii_diag}). In-spiral from stellar tides is responsible for the $\gtrsim 300 M_\oplus$  Jovians inside $2 r_{\rm loss}$, see text for discussion. Observed planet masses and orbital periods with F/G/K hosts ($0.6 M_\odot$ to $1.4 M_\odot$) are denoted by black dots, with the corresponding semi-major axis assuming a solar host displayed on top.
  }
  \label{fig:mass_period_plot}
\end{figure*}

While the $f$-mode response depends strongly on the distance of close approaches, how eccentricities are excited can affect circularization by dynamical tides. \cite{Wu2018} suggested stochastic tides might decouple a gas giant from the secular forcing from a companion planet or star, saving the gas giant from tidal disruption, explaining the observed clustering of hot Jupiter orbits. Later works confirmed this to be the case, provided that the orbit decay timescale from dynamical tides $t_{\rm decay} \sim P_{\rm orb} |E_{\rm orb}/\Delta E|$ is shorter than the timescale orbits are driven to high eccentricities \citep{Vick+2019, Teyssandier+2019}. Recently, \cite{Liveoak+2026b} reaffirmed the \cite{Wu2018} finding that hot Jupiters that reside in the pile-up could have circularized by stochastic tides.

We expand upon this suggestion, with shock-driven outflows setting the periastron separation where gas envelopes become unbound. We presume most hot Jupiters and sub-Saturns are assembled at larger distances, where solid cores that support gaseous envelopes can grow to $\sim 10 M_\oplus$.
\citep[e.g.][]{Pollack+1996, Lee+2014, Lee2019}. Typically, the mechanism that excites orbits to high-eccentricities acts {\it slowly}, over timescales longer than $t_{\rm decay}$, such as Lidov-Kozai oscillations \citep[e.g.][]{Vick+2019} or secular chaos \citep[e.g.][]{Teyssandier+2019}.  In this case, $r_{\rm peri}$ shrinks faster than $a$, until $r_{\rm peri} < r_{\rm circ}$, where stochastic tides decouples the orbit from the perturber. Rapid circularization from dynamical tides halts eccentricity excitation, freezing $r_{\rm peri}$ near $r_{\rm circ}$, saving hot Jupiters from tidal outflows or engulfment by their hosts. Shocks from $f$-modes continue to dwindle orbital energy until the decreasing $a$ dips $r_{\rm peri}$ below $r_{\rm circ}$ (when $\omega \Delta P_{\rm orb}<1$). Stochastic growth of $f$-modes no longer dominates circularization, which continues due to a tidal dissipation mechanism with a higher quality factor $Q$, such as inertial waves \citep[e.g.][]{Dewberry2024, Dewberry2026}, viscosity from turbulent convection \citep[e.g.][]{Duguid+2020, Terquem+2026}, or core visco-elasticity \citep[e.g.][]{StorchLai2014, StorchLai2015}.  Because circularization via planetary tides roughly conserves orbital angular momentum, planets are placed at the final $a \approx 2 r_{\rm peri}$ (Fig.~\ref{fig:HJ_vs_DN_formation} left panel). 

However, some gaseous planets have their orbits elongated \textit{rapidly}, over timescales shorter than $t_{\rm decay}$. Their $r_{\rm peri}$ quickly decrease below $r_{\rm circ}$ until $r_{\rm peri} < r_{\rm loss}$, and close approaches strip gas every orbit. While the planet circularizes from dynamical tides, shock-driven outflows transform a cold Jupiter into a sub-Saturn within the Neptune desert (``desert Neptune'' or ``desert dweller''), migrating to $a \approx 2 r_{\rm peri}$ (Fig.~\ref{fig:HJ_vs_DN_formation} right panel). 

If $f$-modes are both the saviors and destroyers of hot gaseous worlds, we expect the remnant cores of giant planets to lie interior to $2 r_{\rm loss}$, and orbits to cluster between semi-major axes of $2 r_{\rm loss}$ and $2 r_{\rm circ}$. Below masses of $\lesssim 300 M_\oplus \approx 1 M_{\rm Jup}$, our predictions generally agree with the period distributions of hot Jupiters and Neptunes (Fig.~\ref{fig:mass_period_plot}). The boundary of the Neptune desert seems to be well-traced by $2 r_{\rm loss}$. The hot-Jupiter pileup and Neptune ridge lies between $2 r_{\rm loss}$ and $2 r_{\rm circ}$, where $r_{\rm circ}$ is evaluated assuming the semi-major axis is 10 AU before stochastic tides circularize the orbit. A distribution of clustered periods could reflect a range of $2r_{\rm circ}$ values, provided the semi-major axis at formation lies between $\sim$0.1~AU and $\sim$10~AU (Fig.~\ref{fig:radii_diag}). 

Many super-Jupiters ($\gtrsim 300 M_\oplus \approx 1 \Mjup$) bleed interior to $2 r_{\rm loss}$ , likely due to dissipative tides in the host causing inward migration \citep[e.g.][]{Jackson+2008, OwenLai2018, BaileyBatygin2018}. Recently, \cite{Millholland+2025} showed tidal in-spiral predicts the hot Jupiter period distribution behaves like a power-law. The theory of resonance locking, where resonantly-excited stellar gravity-modes lock and cause orbital decay as the host evolves along the main-sequence, give occurrence rates in agreement with those observed.  However, the planet mass below which resonances don't lock depends sensitively on the non-linear dissipation of gravity-modes in stellar cores, which remains highly uncertain \citep[e.g.][]{MaFuller2021, Zanazzi+2024}. We defer the inclusion of tidal in-spiral to future works.

\subsection{Removing Residual Envelopes with Photoevaporation}
\label{sec:Photoevap}

Oscillations proceeding periastron passages no longer drive outflows when the envelope mass becomes less than that of the core ($M_{\rm env} \lesssim M_{\rm core}$), and the bulk density becomes small ($\lesssim 1 \ {\rm g}/{\rm cm}^3$, Fig.~\ref{fig:mesa_gyre_plot}). Such puffy planets lie in sharp contrast to the bulk densities of desert dwelling Neptunes, which span the range $\sim$1-10 ${\rm g}/{\rm cm}^3$ (Fig.~\ref{fig:subSat_density}). Mass loss driven by Roche-lobe overflow shares this conundrum \citep[e.g. Fig.~4 of][]{HallattMillholland2025}, because tidal forcing of any kind struggles to strip dense bodies. However, photoevaporation might ameliorate the under-dense sub-Saturns produced by periapsis shocks. If former hot Jupiters shed the majority of their mass through surface gravity wave breaking, a newly-formed hot Neptune could remove a residual envelope if circularization takes less than $\lesssim$$10^8$-$10^9$ years, when the host spectrum is hard enough to drive a photoevaporative wind. 

We briefly explore how photoevaporation might alter the fates of sub-Saturns that find themselves at sub-three-day orbits, after $f$-mode shocks have removed most mass from a former long-period Jovian. 
Gas giants form with initial semi-major axis of $5 \ {\rm AU}$, whose eccentric orbits excite $f$-modes in the Outflow regime, and evolve their masses, eccentricities, and semi-major axes according to the iterative map described in Section~\ref{sec:LongTerm} for 0.5~Myr (Fig.~\ref{fig:orb_ev}). Planets are deposited at $a = 2 r_{\rm peri}$ after a higher-$Q$ tidal process damps the remaining eccentricity, varying the stellar age corresponding to the time the orbit circularizes $t_{\rm circ}$ between $10^8 - 10^9$~years. Any radius inflation caused by higher-$Q$ circularization is neglected for simplicity, which could inflate radii of Jovians \citep[e.g.][]{IbguiBurrows2009, Leconte+2010, Glanz+2022} and Neptunes \citep[e.g.][]{Millholland+2020, YuDai2024, HallattMillholland2026b} at planetary ages $\lesssim 1$~Gyr.
The in-spiral of Neptunes with $\lesssim 1 \ {\rm day}$ orbital periods is also neglected \citep{LeeOwen2025}, which will decrease $a$ below $2 r_{\rm peri}$ at stellar ages larger than $\gtrsim 1 \ {\rm Gyr}$.

We then see if photoevaporation can remove the residual gas atop these $10 M_\oplus$ sub-Saturn cores. Our estimates adopt the crude energy-limited approximation for X-ray and Ultraviolet (XUV) driven mass loss,
\begin{equation}
    \frac{G M_p}{R_p} \frac{\der M_p}{\der t} = \frac{\eta_{\rm XUV}}{4} L_{\rm XUV} \left( \frac{R_p}{a} \right)^2,
    \label{eq:dotM_photo}
\end{equation}
where $\eta_{\rm XUV}$ is the dimensionless efficiency of incident luminosity from XUV irradiation by the host star at unbinding mass. Many works have shown $\eta_{\rm XUV}$ is not constant, but varies depending on the incident XUV flux, photospheric temperature, and surface gravity \citep[e.g.][]{MurrayClay+2009, OwenJackson2012, Salz+2016, Caldiroli+2022, Broome+2025}. We use the \cite{OwenWu2017} scaling of $\eta_{\rm XUV}$ with escape velocity, valid for nearly Neptune-mass planets:
\begin{equation}
    \eta_{\rm XUV} \approx \frac{0.1}{K_{\rm RL}} \left( \frac{15 \ {\rm km}/{\rm s}}{v_{\rm esc}} \right)^{2},
    \label{eq:eta_photo}
\end{equation}
where
\begin{align}
    &K_{\rm RL} = \frac{(\lambda - 1)^2(2 \lambda + 1)}{2 \lambda^3}, \\
    &\lambda = \left( \frac{M_p}{3 M_\star} \right)^{1/3} \frac{a}{R_p} \sim \frac{R_{\rm RL}}{R_p}
\end{align}
encodes the extra efficiency of photoevaporation when the Roche radius approaches that of the planet \citep{Erkaev+2007}. As in \cite{LeeOwen2025}, we assume that a sun-like host has XUV irradiation that falls like
\begin{equation}
    L_{\rm XUV} = 10^{-3.6} L_\odot \left( \frac{100 \ {\rm Myr}}{t} \right)^{0.86}
    \label{eq:L_XUV}
\end{equation}
at ages $t \ge 100 \ {\rm Myr}$. Equation~\eqref{eq:L_XUV} accounts for the slower decay of extreme ultraviolet irradiation relative to X-ray, which dominates the cumulative irradiation energy at late times \citep[e.g.][]{KingWheatley2021}.

Unlike outflows driven by $f$-modes, photoevaporation strips mass over timescales comparable to the planet's Kelvin-Helmholtz time. The envelope radius $R_{\rm env} = R_p - R_{\rm core}$ shrinks with age as
\begin{equation}
    R_{\rm env}(t) \approx R_{\rm env, 1Gyr} \left( \frac{1 \ {\rm Gyr}}{t} \right)^{0.11}.
\end{equation}
\cite{LopezFortney2014} found the solar composition envelope decline $R_{\rm env} \propto t^{-0.11}$ varies little with planet mass and host irradiation past $10^8$ years. We lift tabulated envelope radii for a 1 Gyr old planet $R_{\rm env, 1Gyr}$ from our \texttt{MESA} models (Fig.~\ref{fig:mesa_gyre_plot}), and linearly interpolate between grid points. Planet radii with age are then computed with $R_p(t) = R_{\rm core} + R_{\rm env}(t)$.


\begin{figure}[htbp]
  \centering
  \includegraphics[width=\columnwidth]{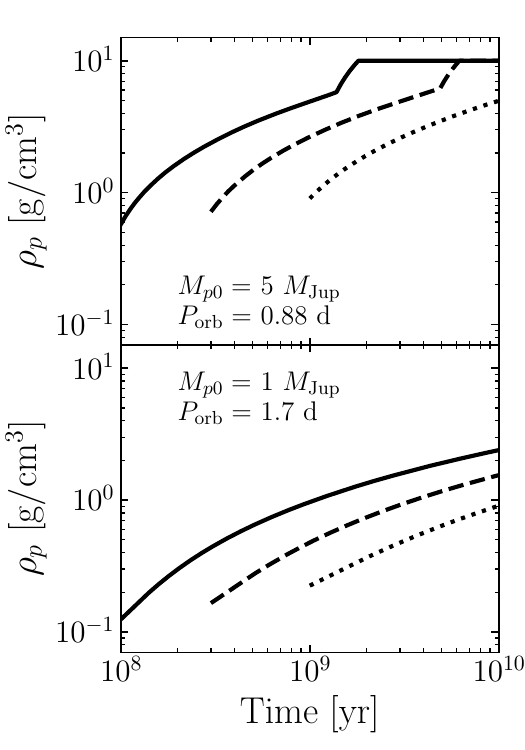}
  \caption{After $f$-mode shocks unbind gas giant envelopes, hot Neptunes arrive in the desert with more gas than that implied by their observed high densities (Fig.~\ref{fig:subSat_density}), which photoevaporation may struggle to remove. Photoevaporation calculations place planets at semi-major axes $a = 2 r_{\rm peri}$ after circularizing at ages $t_{\rm circ} = 10^8, 3\times 10^8, 10^9$ years (solid, dashed, and dotted lines, respectively). Planet masses at $t_{\rm circ}$ are lifted from our iterative map for gas-giants that suffer mass loss, for $\{M_{p0}, r_{\rm peri}\}$ values of $\{5 M_{\rm Jup}, 0.009 \ {\rm AU}\}$ (top) and $\{1 M_{\rm Jup}, 0.014 \ {\rm AU}\}$ (bottom). Cores have $10 M_\oplus$ masses with $10 \ {\rm g}/{\rm cm}^3$ densities.
  }
  \label{fig:photoevap}
\end{figure}

We find the ability of photoevaporation to remove gas depends strongly on orbital period. At periods longer than $\sim$ 1-2 days, bulk densities of our sub-Saturn remnants struggle to climb above $\gtrsim 4 \ {\rm g}/{\rm cm}^3$ (Fig.~\ref{fig:photoevap} bottom panel). Larger $r_{\rm peri}$ separations excite smaller oscillations, unbinding less of the planet's envelope, leaving several Earth masses worth of gas atop the core. Even when circularization places a $14 M_\oplus$ mass sub-Saturn at 1.7 days in $10^8$ years, much of the envelope mass is retained after several Gyr. On the other hand, Neptunes at periods $\lesssim$1 day have little trouble loosing their envelopes. Shorter periastrons lead to stronger shocks that unbind more mass, forming a 0.9 day, $11 M_\oplus$ sub-Neptune after circularizing; only when circularization is delayed ($t_{\rm circ} \gtrsim 1 \ {\rm Gyr}$) is gas retained after several Gyr (Fig.~\ref{fig:photoevap} top panel). 

All calculations carried out in this work set solar composition gas envelopes on top of $10 M_\oplus$ cores. However, some dense $\gtrsim$4 ${\rm g}/{\rm cm}^3$ desert Neptunes have masses larger than $\gtrsim$40  $M_\oplus$  (Fig.~\ref{fig:subSat_density}), requiring larger cores or metal-rich envelopes to reproduce their properties \citep[e.g.][]{Thorngren+2016, Chachan+2025}. For a $\gtrsim$40 $M_\oplus$ core, a solar composition envelope has a $\sim$1 ${\rm g}/{\rm cm}^3$ bulk density when the gas-to-core mass ratio is of order unity. Our preliminary calculations find that wave breaking delivers sub-Saturn with $\gtrsim 40 M_\oplus$ cores into the desert with $\sim$1 ${\rm g}/{\rm cm}^3$ bulk densities. However, these same sub-Saturns also have large escape velocities, which decreases $\eta_{\rm XUV}$ and mutes the ability of photoevaporation to remove residual envelopes, with bulk densities never climbing above $\gtrsim$4 ${\rm g}/{\rm cm}^3$ after several Gyr. Metal rich envelopes encounter similar setbacks, since high metallicities cause a strong drop in $\eta_{\rm XUV}$ \citep[e.g.][]{OwenJackson2012, OwenMurray-Clay2018}. Hydrodynamical simulations with radiative diffusion and pressure, whose planets have massive cores and metal rich envelopes, are needed to ascertain if $f$-mode shocks can produce $\gtrsim$40 $M_\oplus$ sub-Saturns with $\gtrsim$4 ${\rm g}/{\rm cm}^3$ bulk densities (see discussion around eqs.~\ref{eq:zeta_wb}-\ref{eq:zeta_p}).

\section{Summary and Discussion}
\label{sec:SummDisc}

\subsection{Summary}

We argue that fundamental-mode ($f$-mode) oscillations play a crucial role in high-eccentricity migration, setting the occurrence rates of gaseous exoplanets that arrive at short separations after forming at several AU. Our hydrodynamical simulations of gas giants on eccentric orbits tidally forced by their host stars show surface gravity waves ($f$-modes) dissipate through shocks after close approaches (Sec.~\ref{sec:hydro_sim}). Surface gravity wave breaking heats atmospheres, and can drive an outflow when waves break deep. The magnitude and rate that energy is deposited by tidal work, and mass entrained in the shock-heated envelope, from our simulations lie in agreement with normal mode theory. We investigate the structural and orbital evolution of planets that circularize through stochastic tides with an iterative map that utilizes normal mode scalings (Sec.~\ref{sec:Outcomes}). The stochastic growth of $f$-modes circularize orbits over $\lesssim 10^6$ year timescales, with mode energies limited by dissipation from shocks. Shocks strip mass only when the periastron distance is sufficiently short, powering a super-Eddington wind. Shallow shocks that cool by radiative diffusion retain the planet's mass, but damp eccentricities through stochastic tides, clustering sub-Saturn and Jovian planets into the Neptune ridge and hot-Jupiter pile-up. When eccentricity excitation is strong, the orbit is unable to be circularized by stochastic tides, and deep periastron plunges lead to super-Eddington winds powered by wave-breaking. Outflows strip gas giants of their envelopes, placing sub-Saturn cores in the Neptune desert (Fig.~\ref{fig:HJ_vs_DN_formation}).

\subsection{Theoretical Shortcomings}

Although our iterative map shows that deep periastron passages can drive winds that unbind most of the planetary envelope, Neptune-mass cores placed in the desert retain a problematic amount of gas. Maps halt mass loss when gas-to-core mass ratios are of order $M_{\rm env}/M_{\rm core} \sim 0.1 - 1.0$, placing desert dwellers at sub-three-day orbits with bulk densities $\lesssim 1 \ {\rm g}/{\rm cm}^3$, much smaller than those observed (Fig.~\ref{fig:subSat_density}). Photoevaporation may not be able to remove residual envelopes if circularization places sub-Neptune cores past $\sim$1-2 day orbital periods (Sec.~\ref{sec:Photoevap}). A resolution may involve improving our mass loss prescriptions for sub-Saturns and sub-Neptunes.
Most gas giant mass is lost in as little as $\sim 10 - 100$ orbits, and outflows are just as rapidly quenched once gas pressure supports envelopes (see discussion around eqs.~\ref{eq:zeta_wb}-\ref{eq:zeta_p}). 3D simulations or calculations that track mass loss over multiple orbits, include radiative cooling that quenches outflows, and the transition from gas to radiation pressure once shocks deposit energy at super-Eddington rates, are needed to determine the final densities of desert Neptunes.

We reaffirm the findings of \cite{Mardling1995b, VickLai2018, Wu2018} that stochastic tides circularizes fast but dies young, halting abruptly once the mode phase change between successive periastron passages is smaller than unity. Mode amplitudes stagnate, freezing the semi-major axis past $a \gtrsim 0.1 \ {\rm AU}$ at eccentricities larger than $e \gtrsim 0.8$ (Fig.~\ref{fig:orb_ev}). Tidal dissipation must continue through some other process to damp remaining eccentricities.

Wave breaking sets in at displacement amplitudes $|\xi_r/R_p| \sim q_{\rm wb} \sim 10^{-2}$ (Fig.~\ref{fig:mesa_gyre_plot}). This occurs before mode-mode couplings can affect mode growth, which are less effective at limiting mode amplitudes \citep[e.g.][]{Mardling1995b}. Shocks not only damp $f$-modes, but also $p$-modes, and whether wave breaking will permit $p$-modes to non-linearly couple and de-tune $f$-modes is unclear \citep{Yu+2021, Yu+2022}. Shocks occur well before the mode self-gravity causes a hydrodynamic instability at $q \sim 10^{-1}$ \citep[e.g.][]{Yu+2026}, which could save gas giants from tidal disruption.

Our hydrodynamical simulations and iterative map neglects back-reaction torques from mass-transfer, which can modify the orbit. Analytic arguments \citep[e.g.][]{Sepinsky+2007, Weldon+2026a, Weldon+2026b} and hydrodynamical simulations \citep[e.g.][]{Guillochon+2011, Liu+2013} suggest this back-reaction could be large, sometimes unbinding the planet orbit for the deepest periastron passages. \cite{FanLiu2026} argue that when gas giants have large cores, the mass-loss process is stabilized by the binding energy of the core. Including orbit back-reaction might unravel why some giant planet cores reside in the desert, and others become unbound.

\subsection{Observational Predictions}

The large densities and metal rich host stars of desert Neptunes suggests that they are denuded cores of failed hot Jupiters \citep{VissapragadaBehmard2025}. We argue that during high-eccentricity migration (HEM), surface gravity wave breaking unbinds the gaseous envelopes, placing Jovian cores at sub-three-day orbital periods. 
The competing hypothesis Roche-Lobe Overflow (RLO) theorizes desert dwellers are produced after gas giants in-spiral so close to their hosts, stellar tides strip their envelopes \citep[e.g.][]{Lai+2010, Valsecchi+2014, Thorngren+2023, HallattMillholland2025}. Observations can distinguish desert dweller arrival by HEM or RLO:
\begin{enumerate}

    \item HEM expects spin-orbit misalignments for desert Neptunes, RLO expects spin-orbit alignment. Migration via stellar tides not only decreases hot Jupiter separations, but damps stellar obliquities \citep[e.g.][]{BarkerOgilvie2009, Winn+2010, Zanazzi+2024, ZanazziChiang2025}. HEM creates desert dwellers early ($\lesssim 1$ Gyr), before in-spiral damps obliquities. RLO follows period decay caused by stellar tides, producing aligned desert dwellers.
    
    \item HEM predicts a larger age dispersion than RLO. HEM can take as little as $\sim$100~Myrs to place a sub-Saturn core in the desert, while RLO takes several~Gyrs. Both HEM and RLO expect desert dwellers to be older than hot Jupiters that tidally in-spiral, supported by age estimates from galactic kinematics \citep[e.g.][]{HamerSchlaufman2019, HallattMillholland2025}.
    
    \item Stochastic tides during HEM can halt eccentricity growth before mass loss, explaining the excess of hot-Jupiters in the 3-5 day pile-up, Neptunes in the 3-6 day ridge, and deficit of Neptunes at sub-three-day periods \citep{Castro-Gonzalez2024a}. If most hot Jupiters in-spiral \citep[e.g.][]{Millholland+2025}, why RLO produces desert dwellers with small occurrence rates is less obvious.
    
    \item Desert dwellers with close companions could have less tumultuous histories, more easily reconcilable with RLO. 
    Only small planets interior to $\lesssim$4 days become de-stabilized during RLO, which occurs during hot Jupiter in-spiral \citep{Livveoak+2026a}. On the other hand, HEM must relocate gas giants from $\gtrsim$1 AU to $\lesssim$0.1 AU, which can de-stabilize small companions over a much wider range of orbital periods \citep[e.g.][]{Mustill+2015, Becker2026}. 
    
    \item Wide companions to desert dwellers could facilitate dynamical instability, consistent with HEM. The fraction of hot Jupiters that have wide and massive companions ($\gtrsim 1 \ \Mjup$, $\gtrsim 1 \ {\rm AU}$) is $\sim$50\% (e.g.~\citealt{Knutson+2014, Bryan+2016, ZinkHoward2023}). This high value supports a turbulent history.  RLO of hot Jupiters occurs with or without companions, and predicts a similar wide companion fraction. HEM can be catalyzed by massive perturbers, and expects a wide companion fraction $\gtrsim$50\%.
    
    \item Both HEM and RLO struggle to produce high-density desert Neptunes (see e.g. Sec.~\ref{sec:Photoevap}, Fig.~4 of \citealt{HallattMillholland2025}), because tidal excitation and stripping is more difficult for compact bodies. For both HEM and RLO, time-dependent calculations of outflows and radiative cooling are needed to fully access if the predicted desert dwellers are systematically under-dense.
    
\end{enumerate}


Proto-hot Jupiter $f$-mode shocks should also cause bright transients, potentially detectable by the Nancy Grace Roman Space Telescope and Vera C. Rubin Observatory. Our preliminary estimates find that the luminosity of tidal flares can rival and exceed a solar host, heating the atmosphere to temperatures of order $\sim 10^3$-$10^4$~K. Recently, \cite{Kenworthy+2023} discovered an optical and infrared transient around the young ($\sim$300 Myr) solar-like star ASASSN-21qj, that peaked at the blackbody temperature of 1000 K, increasing the system's luminosity by 4\%. Although interpreted as a planetary collision, the ASSASSN-21qj transient could be consistent with a gas giant periapsis shock. Work on flare light-curves from eccentric gas giant periapsis shocks is currently ongoing.

\vspace{0.2in}
\noindent 
We thank Sharone Abhilash, Eugene Chiang, Timothy Hallatt, Sarah Millholland, Ritika Sethi, Michelle Vick, and Shreyas Vissapragada for useful conversations. JJZ was supported by a 51 Pegasi b fellowship from the Heising-Simons Foundation. The Center for Exoplanets and Habitable Worlds is supported by the Pennsylvania State University and the Eberly College of Science. 

\texttt{MESA} inlists are available on Zenodo \citep{zenodo}.


\vspace{5mm}

\facilities{NASA Exoplanet Archive \citep{Christiansen+2025}}

\software{astropy \citep{Astropy_1,Astropy_2,Astropy_3},
          Athena++ \citep{Stone+2020},
          GYRE \citep{TownsendTeitler2013, TownsendZweibel2018, GoldsteinTownsend2020}
          MESA \citep{Paxton+2011,Paxton+2013,Paxton+2015,Paxton+2018,Paxton+2019,Jermyn+2023}, 
          numpy \citep{numpy_cite},
          pandas \citep{pandas_cite},
          scipy \citep{scipy_cite}
          }
\appendix

\section{Higher Periastron Simulations}
\label{sec:wide_peri}

\begin{figure*}
  \centering
  \includegraphics[width=\linewidth]{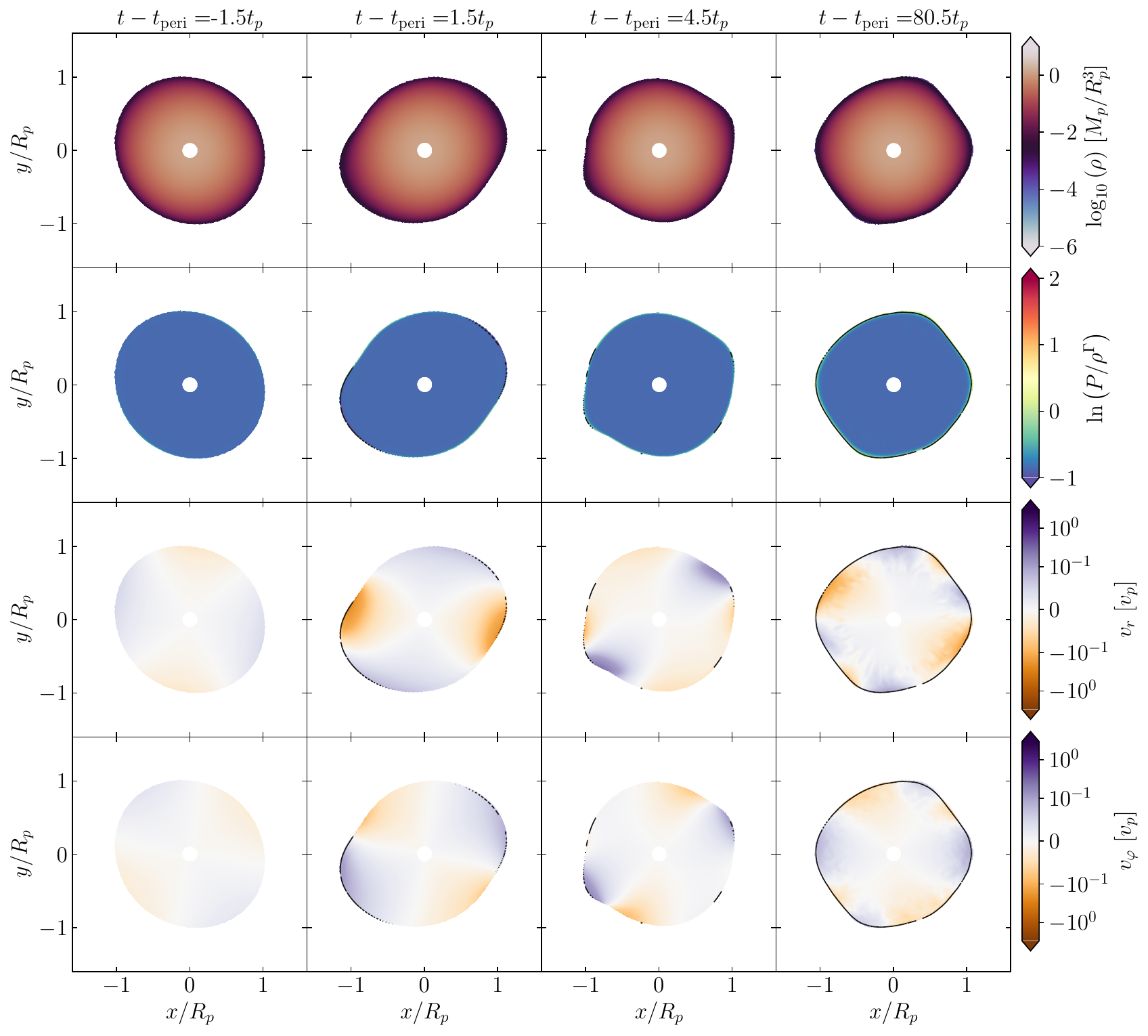}
  \caption{
  Although simulations with wider periastron passages excite tidal flows, they drive small or non-existent shocks that don't power outflows. Colors, lines, and labels for all panels have meanings identical to Fig.~\ref{fig:peri_pass_profiles}, except $r_{\rm peri} = 2.4 r_t$.
  }
  \label{fig:peri_pass_profiles_largeperi}
\end{figure*}

\begin{figure}
  \centering
  \includegraphics[width=\columnwidth]{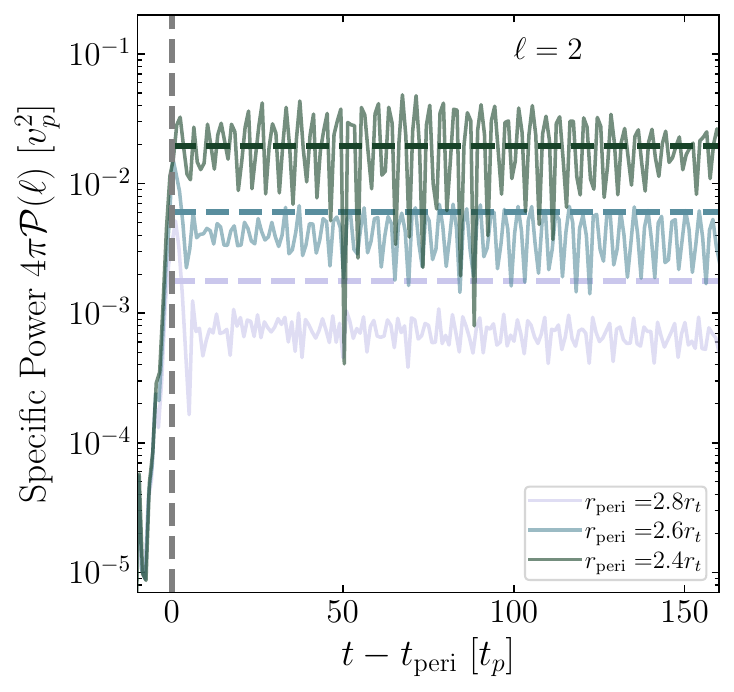}
  \caption{ 
  Higher periastron passages have little power decay. Thin solid lines denote $\ell=2$ specific power for $r_{\rm peri}$ values indicated, while thick dashed lines display linear theory predictions, assuming $\ell=2$ power remains constant after impulsive excitation at periastron (eq.~\ref{eq:vlm_linear}). Vertical dashed line denotes $t = t_{\rm peri}$. Exponential decay fits to $\mathcal{P}(2)$ (eq.~\ref{eq:P2_fit}) give $\gamma = \{1.3 \times 10^{-3}, 8.2 \times 10^{-4}, 4.4 \times 10^{-4}\} t_p^{-1}$ for $r_{\rm peri} = \{2.8, 2.6, 2.4\} r_t$.
  }
  \label{fig:power_time_largeperi}
\end{figure}

Because distant periastron passages excite weak tidal flows that do not shock, we use distant encounters to verify wave breaking occurs when radial velocities are super-sonic, as well as constrain the numerical dissipation in our simulation grid. Weak shocks are expected for $r_{\rm peri} = 2.4 r_t$, since linear theory finds the $f$-mode radial velocity slightly exceeds the sound-speed ($v_r|_{\rm max} \approx 0.11 v_p> c_{\rm s}|_{\rm min} \approx 0.06 v_p$). Our $r_{\rm peri} = 2.4 r_t$ simulation confirms shocks occur in a thin veneer near the planet surface ($\ln[P/\rho^\Gamma] \approx 0$ near $r\approx R_p$, Fig.~\ref{fig:peri_pass_profiles_largeperi}, $2^{\rm nd}$ row, $4^{\rm th}$ column), and the outflow density never exceeds that of the artificial background (Fig.~\ref{fig:peri_pass_profiles_largeperi} top row). For $r_{\rm peri} = 2.6 r_t$, no shocks are seen in our simulation, in agreement with theory since $v_r|_{\rm max} < c_{\rm s}|_{\rm min}$.

If dissipation from shocks is the cause of $\ell=2$ power decay for close encounters, wide encounter $\ell=2$ power should remain approximately constant. Figure~\ref{fig:power_time_largeperi} displays a set of large $r_{\rm peri}$ simulations, with either weak ($r_{\rm peri} = 2.4 r_t$) or no ($r_{\rm peri} = \{2.6, 2.8\} r_t$) shocks. Encounters with $r_{\rm peri} = 2.8 r_t$ are the weakest we can simulate, because more distant pericenters give radial displacements smaller than our grid cells. Not only does linear theory agree with the simulated $\ell=2$ power within a factor of $\sim$3, the amplitude of the flow does not appreciably decay with time, with exponential decay fits finding $\gamma$ values between $4.4 \times 10^{-4} - 1.3 \times 10^{-3} t_p^{-1}$. As expected, closer encounters $(r_{\rm peri} = \{1.8, 2.0\} r_t)$ display a steeper decay in power (Fig.~\ref{fig:power_spectrum}), due to physical dissipation from shocks. However, although our $r_{\rm peri} = 2.2 r_t$ simulation clearly displays shocks (Fig.~\ref{fig:peri_pass_profiles}), we caution that inferred damping rate lies near the noise limit of our grid, and the measured $\gamma$ value could be inaccurate because of inadequate grid resolution, or interaction with the artificial background.

\bibliography{main}{}
\bibliographystyle{aasjournal}



\end{document}